\documentclass[secnumarabic, graphics,floatfix, nofootinbib,tightenlines,nobibnotes, aps, prl, 12pt]{revtex4-1}
\usepackage{graphicx}
\usepackage{amsmath}
\usepackage{natbib}

\begin{document}

\title{Decomposition of fluctuating initial conditions and flow harmonics}
\author{Wei-Liang Qian$^{1,2}$, Philipe Mota$^3$, Rone Andrade$^1$, Fernando Gardim$^1$, Fr\'ed\'erique Grassi$^1$, Yogiro Hama$^1$, Takeshi Kodama$^4$}
\affiliation{$^1$Universidade de S\~ao Paulo, SP, Brazil}
\affiliation{$^2$Universidade Federal de Ouro Preto, MG, Brazil}
\affiliation{$^3$Frankfurt Institute for Advanced Studies, Frankfurt am Main, Germany}
\affiliation{$^4$Universidade Federal do Rio de Janeiro, RJ, Brazil}

\date{October, 19, 2013}

\begin{abstract}
Collective flow observed in heavy ion collisions is largely attributed to initial geometrical fluctuations,
and it is the hydrodynamic evolution of the system that transforms those initial spatial irregularities into final state momentum anisotropies.
Cumulant analysis provides a mathematical tool to decompose those initial fluctuations in terms of radial and azimuthal components.
It is usually thought that a specified order of azimuthal cumulant, for the most part, linearly produces flow harmonic of the same order.
In this work, by considering the most central collisions (0-5\%), we carry out a systematic study on the connection between cumulants
and flow harmonics using a hydrodynamic code called NeXSPheRIO.
We conduct three types of calculations,
by explicitly decomposing the initial conditions into components corresponding to a given eccentricity
and studying the out-coming flow through hydrodynamic evolution.
It is found that for initial conditions deviating significantly from Gaussian, such as those from NeXuS,
the linearity between eccentricities and flow harmonics partially breaks down.
Combining with the effect of coupling between cumulants of different orders,
it causes the production of extra flow harmonics of higher orders.
We argue that these results can be seen as a natural consequence of the non-linear nature of hydrodynamics,
and they can be understood intuitively in terms of the peripheral-tube model.
\end{abstract}

\maketitle
\newpage

\section{I. Introduction}

Event-by-event initial-geometry fluctuations have proved to be an essential ingredient in ongoing efforts to study the collective
flow observables and to describe specific features of particle correlations \cite{sph-1st,sph-v2-1,sph-v2-2,hydro-eve-1,hydro-eve-2,sph-corr-1,sph-corr-3,sph-corr-2,hydro-v3-1,hydro-v3-3,hydro-v3-5,hydro-eve-3,hydro-eve-4,epos-2,hydro-eve-5,hydro-eve-6,hydro-eve-8,hydro-eve-9},
measured at Brookhaven National Laboratory's Relativistic Heavy-Ion
Collider (RHIC) \cite{star-ridge2,star-ridge3,phenix-ridge5,phobos-ridge6,phobos-ridge7}
and CERN's Large Hadron Collider(LHC) \cite{cms-vn1,cms-vn2,cms-ridge3,cms-ridge4,alice-vn1,alice-vn2,alice-vn3,alice-vn4,atlas-vn1,atlas-vn2,atlas-vn3,atlas-vn4}.
Recently, much efforts have been devoted to investigate how the hot spots in the initial conditions (IC) of heavy-ion collisions
may translate into the observed harmonic components $v_n$ of the flow \cite{hydro-v3-1,hydro-v3-3,sph-vn-2,sph-vn-3,hydro-vn-1,hydro-vn-2,ph-vn-3,ph-vn-6,hydro-vn-3,hydro-vn-4}.
In particular, the third component of Fourier expansion, triangular flow ($v_3$),
has received much attention \cite{hydro-eve-3,hydro-v3-2,hydro-v3-4,hydro-v3-5,hydro-v3-6,hydro-v3-7,hydro-eve-6}.
The focus on triangular flow was in part triggered by the work of Alver and Roland \cite{hydro-v3-1},
where they introduced the concept of triangular flow, caused by triangularity in the fluctuating initial-conditions.
Since the collective flow is understood as the medium response to the event-by-event geometrical fluctuations in IC,
the observed features of the flow could be used to extract information on the initial spatial irregularities of the colliding systems
and transport properties of quark-gluon plasma (QGP).
In \cite{hydro-v3-1}, they argued that the triangular-flow coefficient $v_3$ may capture a significant portion of the observed ``ridge'' and ``shoulder'' structures \cite{star-ridge1,star-ridge2,star-ridge3,phenix-ridge4,phenix-ridge5,phobos-ridge6,phobos-ridge7,alice-vn4,cms-ridge5,atlas-vn1}
in di-hadron azimuthal correlations.
In their picture,
the underlying physics of certain features of two-particle correlation itself,
such as two symmetrical and lower away-side peaks in the central collisions, in-plane/out-of-plane effect in mid-central collisions, etc. 
does not appear transparently.
However, the crucial point of their interpretation is that it is a hydrodynamic explanation, {\it i.e.},
they attribute the observed features of two particle correlation to the medium itself,
as done earlier in \cite{sph-corr-1,sph-corr-3}, where the dynamical mechanism of ridge formation has clearly been shown.

Since then, many efforts have been devoted to explore the connection between the initial eccentricities and flow harmonics.
It was proposed to decompose the initial geometric configurations by means of cumulant expansion \cite{hydro-v3-2}.
The leading order cumulants in azimuthal angle are defined as eccentricities $\varepsilon_n\,$, analogously to the flow harmonics $v_n\,$.
The spirit of cumulant analysis is that one expects approximate linear response of the medium.
It is inferred that hydrodynamic evolution essentially transforms each cumulant component individually into
a corresponding Fourier coefficient in momentum space.
Since the average Glauber distribution is roughly Gaussian, the expansion is basically carried out around a 
Gaussian form\footnote{Here the term ``Gaussian" is used in the same context as that in Ref.\cite{hydro-v3-2}, also see the discussions below Eq.(3)}.
In central collisions, the dipole term and other higher order terms would be sufficiently small,
so the hydrodynamic response to a specified order of cumulants could be linear to a good approximation.
In other words, the cumulant expansion approach expects the overall effect to be the summation of those from each azimuthal cumulant,
while each azimuthal cumulant transforms a certain order of spatial deformation into corresponding anisotropy in momentum space.
Since experimental measurements of the initial eccentricities have not been possible to date,
current studies are mostly relying on model calculations.
In fact, there have long been speculations \cite{ph-vn-1,ph-vn-2,ph-vn-3,ph-vn-4} about
a linear relation between eccentricities $\varepsilon_n$ and flow harmonics $v_n$ of the produced hadrons.
The ratio of $v_n/\varepsilon_n$ was conjectured to be almost constant suffering suppression at moderately big $n$.
These ideas have been explored in the studies of properties of collective flow and particle correlations.
Numerical calculations based on Glauber Monte Carlo models in conjunction with hydrodynamic or transport models
showed an essentially linear relation between average $v_n$ and $\varepsilon_n$ \cite{hydro-v3-1,hydro-v3-3,hydro-eve-3,hydro-v3-2,hydro-v3-4}.

Meanwhile, more realistic simulations of heavy-ion collisions which attempt to faithfully replicate collision events have also been carried out
aiming at reproducing quantitatively the observed data on collective flow and correlations.
In a hybrid transport model named AMPT \cite{ampt-1}, one has found many features of the observed correlations \cite{ampt-2,ampt-3},
although these results were previously attributed to jet-medium interactions.
By using an ideal hydrodynamic model, called NeXSPheRIO \cite{sph-review-1},
it was shown that the main properties of the measured two particle correlation could be reproduced
if event-by-event fluctuating IC were included \cite{sph-corr-1}.
On the contrary, when one switches off fluctuations and uses smoothed IC (average ones), the ``ridge'' and ``shoulder'' disappear.
Therefore, event-by-event fluctuation was identified as one of the key ingredients to produce major features in the data.
In fact, these calculations helped to establish the consensus on hydrodynamical origin of the long-range di-hadron correlation.
It is worthwhile to point out, IC generated by those realistic event generators are not necessarily close to Gaussian.
For example, as for IC generated by NeXuS \cite{nexus-1,nexus-rept}, the deviation from Gaussian is quite big.
Even with respect to the average distribution, the difference can not be treated perturbatively.
As a result, for such systems, the evolution is expected to deviate significantly from linearity.

As a matter of fact, the deviation from the linear hydrodynamic response has caught much attention recently.
Some efforts have been made to study such deviations in amplitude as well as in event plane orientation.
In \cite{hydro-vn-2}, the authors studied the nonlinear response of $v_n$ of the produced hadrons.
For instance, the directed flow of emitted hadrons $v_1$ was found to be related to the second and third cumulants,
and such non-linear response constitutes about 25\% correction in mid-central collisisions.
Their work can be seen as, in part, inspired by the scaling law $v_4/(v_2)^2=\frac{1}{2}$ (for large $p_T$)
which was first studied by Borghini and Ollitrault\cite{ph-vn-7,ph-vn-8}.
Following their line of thought, the hadrons are emitted from the fluid elements whose velocity distribution can be decomposed as
\begin{eqnarray}
u(\phi)=U(1+2V_2 \cos(2\phi)+2V_4\cos(4\phi)\cdots)\ ,
\end{eqnarray}
where $\phi$ is the azimuthal angle of the fluid velocity with respect to the reaction plane.
Then, the resultant octopole flow of the hadrons $v_4$ consists of two contributions,
a linear contribution from the fourth Fourier coefficient of the fluid $V_4$
and a non-linear contribution from the second Fourier coefficient $V_2$. The latter is proportional to $V_2^2$.
Therefore, the coefficient $w_{4(22)}$ defined in ref.\cite{hydro-vn-2} corresponds to the nonlinear response of $v_4$ proportional to $V_2^2$.
If one further assumes that the harmonic coefficient $V_n$ of the fluid responds {\it linearly}
to $\varepsilon_n$,
$w_{4(22)}$ should basically be proportional to $\varepsilon_2^2$.
As shown in the paper, both the linear and the non-linear coefficients turned out to be functions of centrality as well as transverse momentum.
The findings in \cite{hydro-vn-2} generalize the results previsously found in \cite{sph-vn-3}.
In \cite{sph-vn-3,hydro-vn-3}, the authors studied the relation between initial state eccentricities and final state flow harmonics in a event-by-event base, 
and its deviation from a linear hydrodynamic response.
In \cite{hydro-phin-1,glauber-en-1,glauber-en-2}, the correlation between event plane was investigated. 
It was shown that event planes of different orders are in fact correlated.
Such correlations may also be attributed to the correlations already contained in the eccentricities of the fluctuating IC,
by evaluating the correlations between participant planes in the Glauber model \cite{glauber-en-1,glauber-en-2}.

All of the above works have treated the IC as a whole, 
and the deviation from linearity was usually estimated in accordance with some presumed relations.
To understand the problem more transparently, it is interesting to investigate the behaviour of each individual cumulant.
This motivated us to introduce a somewhat different approach.
In this work, we carry out an explicit study on some specific cumulants and their effects on flow harmonics.
We are conscious of the fact that the results actually depend on the specific choice of model and,
due to the non-linear nature of the hydrodynamics, they can be quite complicated.
We will use the NeXSPheRIO code, since it has been successfully employed to study various phenomena in heavy ion
collisions\cite{sph-v2-1,sph-v2-2,sph-cfo-1,sph-hbt-1}, including long-range two-particle correlations \cite{sph-corr-1}.
Besides, IC generated by realistic event generator NeXuS differ very much (as already mentioned) from Gaussian form,
which makes it a good candidate for the purpose of the present study.
To achieve our goal, we generate fluctuating events, perform Fourier decomposition of the IC
and study the hydrodynamic evolution of individual terms called $e_n(r,\phi)$ (each $e_n$ is related to a given symmetry, eccentricity $\varepsilon_n$ and 
eccentricity plane $\Phi_n$, see section II and III below). 
In particular, we study the effect of both cumulants and eccentricity planes.

In order to understand the results, 
we will make use of a simple {\it peripheral-tube model}, which interprets ``ridge'' and ``shoulders" as {\it causally connected structures}. 
They are produced by deflection of the flow, caused by a highly energetic tube located close to the surface of the fluid.
At a first glimpse, the results of this model may be thought to be equivalent to those based on cumulant analysis and triangular flow,
since the resulting flow indeed contains a big portion of the third harmonic.
However, as we proceed to argue, despite its simplicity,
the tube approach provides an intuitive explanation of the calculated flow, in particular, when linearity does not hold.
Moreover, discussions will be dedicated to clarify the underlying difference between the two approaches and possible consequences.

The paper is organized as follows.
In Section II, we briefly review the cumulant analysis and then explain how we devise the IC to accomplish our goal.
Eccentricity plane rotation is introduced for further use.
Three types of verifications are then proposed to investigate the connection between cumulants and flow harmonics.
In section III, we proceed to decompose NeXuS IC to numerically perform the verifications proposed in Section II.
Discussions are given in section IV, where we show that the resulting flow can be
conveniently understood in terms of our peripheral-tube model.
We discuss the underlying physics as well as address the main difference between our approach and the explanation based on triangular flow.

\section{II. Cumulant analysis and eccentricity planes}

Cumulants and eccentricity planes are defined in Ref. \cite{hydro-v3-2}. We can synthesize, in leading order, as
\begin{eqnarray} 
\label{eqen}
  &&\varepsilon_1\,e^{i \Phi_1}
  = -\frac{\langle r^3 e^{i \phi}\rangle}{\langle r^3\rangle},
\\ 
  &&\varepsilon_n\,e^{i n\Phi_n}
  = -\frac{\langle r^n e^{i n\phi}\rangle}{\langle r^n\rangle},
  \quad (n>1)
\nonumber
\end{eqnarray}
or equivalently,
\begin{eqnarray}
\varepsilon_1 &=& \frac{\sqrt{\langle r^3 \cos(\phi)\rangle^2+\langle r^3 \sin(\phi)\rangle^2}}{\langle r^3\rangle}\ ,  \nonumber\\
\Phi_1 &=&  \mathrm{arctan2} \left({\langle r^3 \sin(\phi)\rangle},{\langle r^3 \cos(\phi)\rangle}\right)+\pi\ , \nonumber \\
\varepsilon_n &=& \frac{\sqrt{\langle r^n \cos(n\phi)\rangle^2+\langle r^n \sin(n\phi)\rangle^2}}{\langle r^n\rangle}\ , \label{dcumulants} \\
\Phi_n &=& \frac{1}{n} \mathrm{arctan2} \left({\langle r^n \sin(n\phi)\rangle},{\langle r^n \cos(n\phi)\rangle}\right)+\frac{\pi}{n}, \quad(n > 1)\nonumber
\end{eqnarray}
\noindent where $\varepsilon_n$ is the $n$-th eccentricity and $\Phi_n$ is the corresponding eccentricity plane,
and the coordinate system is shifted to the center of mass of the participating nucleons such that $\langle x\rangle=\langle y\rangle=0$.
It is noted that for $n=1,2,3$, the form of cumulant coincides with that of the moment\cite{hydro-v3-2}.
In this work, we will only focus on the first three lowest order cumulants.
Following the arguments in Ref.\cite{hydro-v3-2}, the most dominant terms from a cumulant expansion give arise to a distribution characterized 
by average squared radius $\langle r^2 \rangle$ with an elliptic eccentricity $\epsilon_2$.
Higher order corrections in this expansion will further correct the distribution away from the above Gaussian form.
For Glauber type model, high order coefficients are small.
However, for event generator such as NeXuS, higher order corrections can be big enough.
In particular, for central collisions, elliptic deformation arise mainly from event by event fluctutions, 
rather than from the initial geometrical shape of the intersection of the two colliding nuclei, so its contribution can be comparable to other high order corrections.
The arguments of linearity employed by other authors may not hold true in this particular case.
In this case the leading term in the energy density expansion is the second order correction of zeroth harmonic $(n=0)$,
which possesses a Gaussian form with average squared radius $\langle r^2 \rangle$.
In Eqs.(\ref{dcumulants}), $\langle \cdots \rangle = \int\cdots e(r,\phi)rdrd\phi/\int e(r,\phi)rdrd\phi$, where $e(r,\phi)$ is the energy density. 

Flow harmonics and the corresponding flow planes are defined by
\begin{eqnarray} 
\label{eqvn}
  &&v_n\,e^{i n\Psi_n}
  = {\langle e^{i n\phi_p}\rangle},
\nonumber
\end{eqnarray}
or equivalently
\begin{eqnarray}
v_n &=& \langle\cos(n(\phi_p-\Psi_n))\rangle\ ,\label{dflowharmonics}\\
\Psi_n &=& \frac{1}{n} \mathrm{arctan2} \left({\langle\sin(n\phi_p)\rangle},{\langle\cos(n\phi_p)\rangle}\right)\ . \nonumber
\end{eqnarray}

\noindent In Eqs.(\ref{dflowharmonics}), $\langle \cdots \rangle $ is an average on particles emitted in one event, with azimuthal angle $\phi_p$. 

It was speculated that in analogy to the case of elliptic flow, when spatial anisotropy is transformed into momentum anisotropy,
flow direction basically points at where the pressure gradient is the largest.
Therefore the following relation between flow plane $\Psi_n$ and eccentricity plane $\Phi_n$ is approximately satisfied
\begin{eqnarray}
\Psi_n \sim \Phi_n\ ,
\end{eqnarray}
and eccentricity planes have not played an essential role in the studies of the magnitude of flow components.
In fact, it can be shown explicitly for future use that a rotation of the eccentricity plane will not affect the value of the eccentricity.
We introduce an eccentricity-plane rotation as follows, if one rotates a specific eccentricity plane
of $\varepsilon_n$ (of order $n$) by an angle $\Phi_n'$, 
then by making use of the notations in Eqs.(\ref{dcumulants}), one has
\begin{eqnarray}
\varepsilon_n &\to& \frac{\sqrt{\langle r^n\cos(n(\phi-\Phi_n'))\rangle^2+\langle r^n\sin(n(\phi-\Phi_n'))\rangle^2}}{\langle r^n\rangle}
= \varepsilon_n \nonumber
\end{eqnarray}
We see that the cumulant component in question (as well as all the other components) remains unchanged by definition.
This implies that $\{\varepsilon_n, \Phi_{n}\}$ are independent quantities, since modifying one eccentricity plane $\Phi_n$ 
does not affect others.
If flow harmonics are mostly determined by the magnitude of cumulants,
eccentricity plane rotation shall not affect much the resulting flow, neither two particle correlations. This will be tested.

In order to quantitatively investigate the connection between cumulants and flow, we propose three different types of calculations.
The basic idea is to manipulate a given event,
and generate certain new IC which give better insight to the one-to-one mapping of cumulants onto flow harmonics.
This is accomplished by explicit extraction, removal or rotation of designated $e_n$.
Firstly, we want to see whether IC composed of only
one certain order of cumulant,
produce exclusively flow harmonic of the same order.
To achieve this, we numerically extract an individual $e_n$, and feed it to the SPheRIO code.
In this way, it is possible to verify if extra flow harmonics of different order are produced through hydrodynamic evolution and,
if anything, to study the magnitudes of such components.
Secondly, the test is carried out other way around;
we devise IC by eliminating one certain cumulant and check if hydrodynamic evolution may still produce flow of the missing order.
The third test verifies whether eccentricity plane rotations modify the resulting flow coefficients,
which is aimed to investigate the coupling between different orders of cumulants.
In the first step, we pick just one random NeXuS event and apply the proposed calculations.
Then we study the general case by doing event-by-event analysis, computing the average over the events at the end.
An individual event may be very different from others due to fluctuations,
a convincing conclusion can only be drawn if the results are based on event-by-event simulations.
Such study is carried out in this work. We calculate the standard deviation of all the quantities,
such as the magnitudes of the produced flow coefficients, from a set of random NeXuS events.

\section{III. Numerical Results of NeXSPheRIO}

Here we implement the methods described above into NeXuS IC.
All calculations concerning manipulation of a single event are carried out on the same randomly chosen NeXuS event of a central collision (0-5\%).
In Fig.1, we plot its energy density distribution at $\eta=0$, and the calculated eccentricities $\varepsilon_n$ and flow coefficients $v_n$ are shown in Fig.2.
Each $e_n$ is extracted, by doing numerical Fourier series expansion in terms of azimuthal angle $\phi\,$,
for each radial variable $r\,$:
\begin{eqnarray}
e^{\mathrm{NeXuS}}(r,\phi)&=& e_0(r)+\sum_{n=1}^\infty 2[e^c_n(r)\cos(n\phi)+e^s_n(r)\sin(n\phi)]\ .\label{enexpansion}\\
&\equiv& e_0(r)+\sum_{n=1}^\infty e_n(r,\phi)
\end{eqnarray}
Using Eq.(\ref{enexpansion}) in Eq.(\ref{dcumulants}), one gets in terms of $e_n^{c,s}(r)$
\begin{eqnarray}
\varepsilon_n &=& \frac{\sqrt{\left(\int dr r^{n+1} e^c_n(r)\right)^2+\left(\int dr r^{n+1} e^s_n(r)\right)^2}}{\int dr r^{n+1} e_0(r)}\ , \label{cumulanten} \\
\Phi_n &=& \frac{1}{n} \mathrm{arctan2} \left({\int dr r^{n+1} e^s_n(r)},{\int dr r^{n+1} e^c_n(r)}\right)+\frac{\pi}{n}\ , \hspace{1.25in}(n\ne 1),\nonumber\\
\varepsilon_1 &=& \frac{\sqrt{\left(\int dr r^4 e^c_1(r)\right)^2+\left(\int dr r^4 e^s_1(r)\right)^2}}{\int dr r^4 e_0(r)}\ ,  \nonumber\\
\Phi_1 &=&  \mathrm{arctan2} \left({\int dr r^4 e^s_1(r)},{\int dr r^4 e^c_1(r)}\right)+\pi\ .
\nonumber
\end{eqnarray}
Since each couple of coefficients $e_n^{(c,s)}$ only contribute to the cumulant $\varepsilon_n$ of the same order $n$,
one may use them to devise IC leading to cumulants of desired orders.
However, before proceeding to study the hydrodynamic evolution of each component individually,
one has to deal with a tricky point.
A Fourier component carries not only positive but also negative part.
The latter implies negative energy density, therefore it does not correspond to any realistic case.
To avoid the problem of negative energy density,
one may add to those $e_n$ an isotropic background, which fills up the negative energy region.
A natural choice is to make use of $e_0$ as the desired background.
In Fig.3, we show the energy density distributions (only positive part) of $e_1$, $e_2$ and $e_3$
and those entirely summed up with the isotropic background $e_0$.
Since one uses $e_0$ of the original event, all the resulting IC on the r.h.s. of Fig.3
automatically possess the same amount of total energy as the original event.
One may expect that the resulting multiplicities from these events will be very similar.
Therefore, it is easier for us to focus on the difference, which essentially appears in the flow harmonics.

In Fig.3, one can see that the energy distributions of the devised NeXuS events as functions of radius are naturally twisted.
Though they do not possess any azimuthal harmonics of higher order,
they do not look like what are usually presented in pictorial illustrations (for instance, Fig.1 of ref.\cite{hydro-v3-2}),
where only the leading order $\varepsilon_n$ are considered.
Let us take $e_0+e_1$ as an example,
the ``dipole" distribution possess a tail which comes from the radius dependence of the Fourier series coefficients.
Owing to the non-linear nature of hydrodynamics,
one may further expect that hydrodynamic evolution might produce $v_2$ and $v_3$ besides $v_1$.
Similarly, IC consisting of $e_2$ and $e_0$ might also introduce $v_4$ in addition to $v_2$,
thought it would not produce $v_3$ due to symmetry.

We feed these IC to our hydrodynamic code SPheRIO, and calculate the flow harmonics.
When implementing the IC in Fig.3 in SPheRIO, a finite number of SPH particles (around 250,000) has to be used
(see, e.g., ref.\cite{sph-review-1} for more detailed information on SPH method).
As a consequence, for modified IC shown in Fig.3, some small extra eccentricities might be generated\footnote{For smoother IC as in Fig.10, the resulting precision is much better as shown in Fig.11.}.
To estimate the precision of our present numerical approach, 
the eccentricities are recalculated afterwards by using their definitions in Eqs.(\ref{dcumulants}).
In Fig.4, we plot the resultant eccentricities of the three devised events and the flow components.
One sees that due to the resolution of the numerical implementation,
some cumulants expected to be zero are not identically zero, though numerically they are small enough (see e.g. $\varepsilon_4$).
As expected, one can see that $v_2$ and $v_3$ are indeed produced from IC composed of  $e_1+e_0$.
It is also found that $v_4$ and a hint of $v_6$ are also produced in the $e_2+e_0$ and $e_3+e_0$ cases respectively,
though their magnitudes are relatively smaller.

We also did calculations on the event-by-event basis.
These are done by generating 150 NeXuS events in the same centrality window,
and implementing the same procedure on every one of them.
The averaged eccentricities and flow harmonics are presented in Fig.5,
where error bars indicate the standard deviations.
One can see clearly that previous results are not only for an isolated case.
In general, IC composed of $e_n$ not only produces $v_n$, but also $v_m$ with smaller magnitude, where $m$ is a multiple of $n$.
In particular, $e_1$ extracted from NeXuS IC does produce sizable $v_2$ and $v_3$.

Now we present the results of the second test.
Fig.6 shows the energy distribution of a devised IC with $e_3$ eliminated from the original event.
Here, $v_3$ might be produced due to two causes.
Firstly, as one has already seen (Figs.4,5), $e_1$ produces $v_3$.
Secondly, the interaction between different cumulant components
(eg. $e_1$ and $e_2$ as discussed in ref.\cite{hydro-vn-2}) might also contribute to $v_3$.
In Fig.7, we show the resulting flow harmonics of the modified IC and those obtained from the original event (Fig.3).
One can clearly see that sizable $v_3$ can indeed be produced from IC with $\varepsilon_3 \sim 0$. 
Due to the numerical precision, the generated $\varepsilon_3$ is not identically zero.
However, we note that 
it is much smaller than the original $\varepsilon_3$, producing nevertheless $v_3$, which is comparable to the original one shown in Fig.2.
This result is further confirmed by event-by-event analysis.

In our third test, the eccentricity planes of the original event are randomized.
To be more specific, each $e_n$ is assigned to a different random rotation $\Phi_n'$ so that the energy density reads
\begin{eqnarray}
e(r,\phi)\rightarrow e(r,\phi) = e_0(r)+\sum_{n=1}2[e^c_n(r)\cos n(\phi-\Phi_n')+e^s_n(r)\sin n(\phi-\Phi_n')]\ .\label{enrotation}
\end{eqnarray}
Two of those generated events are shown in Fig.8.
By repeatedly carrying out the same procedure on the same event but with different random seeds,
one obtains a lot of seemingly ``different'' events,
which will be used to carry out calculations on event-by-event basis.
Since the cumulants of these events are identical,
the deviation among them actually gives the ``uncertainty" when one simply measures the irregularities of IC
in terms of cumulants defined in Eqs.(\ref{dcumulants}).
In particular, if one calculates the flow coefficients and their standard deviations,
the latter will present themselves as flow fluctuation resulting from the same given eccentricities.
The calculated results are shown in Fig.9.
It can be seen that the linearity property of cumulant analysis breaks down for events generator such as NeXuS.
If one carries out the calculations using different events, the above discussion implies that
a portion of flow fluctuations actually comes from fluctuations in eccentricity planes
rather than from those in the magnitude of eccentricities.
It is impossible to distinguish one from the other merely from flow measurement.

\section{IV. Discussions and Conclusions}

Cumulant analysis can be understood as a measure of the IC granularity from a global aspect of view.
Mathematically, it is alway possible to decompose the initial configurations by using Fourier series.
If the average IC distribution is roughly Gaussian and the fluctuations are small, 
the resultant cumulants are small and the hydrodynamic response of the system is expected to be essentially linear.
As a result, the overall momentum anisotropy can be seen as the summation of the individual Fourier components,
each of which corresponding to an initial spatial deformation of the same order.
For simplicity, let us take the triangular flow as an example.
In this picture, the produced triangular flow is seen as mainly caused by the eccentricities obtained from the initial fluctuating distributions,
particularly triangularity.
Since the maxima of a given $e_n$ ($n \ge 2$) on the transverse plane obtained from IC are spatially separated,
the outgoing hadrons from different flow peaks are not causally related. 
The hydrodynamic evolution in this context transforms globally the initial fluctuating configurations into flow,
which is also randomly fluctuating, without a clear causal relation among the produced particles.

In the case where IC contains hot spots (eg. NeXuS), the situation for causality is different.
This can be more easily seen using the {\it peripheral-tube model} that we developped in a different context.
One might argue that one can always carry out Fourier expansion,
so any explanation will simply be able to fit into the picture of cumulant analysis.
However, as we are about to argue, under certain conditions peripheral-tube model provides a more intuitive explanation,
and some subtle difference exists.

The peripheral-tube model aims at giving an intuitive interpretation of
how various structures in the two particle correlations are generated.
Details of the model can be found in Refs.\cite{sph-corr-3,sph-corr-2,sph-corr-4,sph-corr-5,sph-corr-6,sph-corr-7}.
Here we only summarize the main points.
We noticed that the energy profile on transverse plane can be seen as composed of hot spots
randomly distributed on top of a rather smooth background.
The latter can be obtained by averaging over various events.
It is important to point out that in the case of NeXuS events,
the background distribution is rather edgy and can not be fitted well by using Gaussian parameterization\footnote{Numerically,
a tube placed on a Gaussian background will produce much less flow due to the big Gaussian tail,
comparing to our present approach.\cite{sph-corr-3}}.
Since the energy distribution possesses an approximate boost invariance,
the hot spots are actually high-energy tubes extending in longitudinal direction.
It was shown \cite{sph-corr-4} that the effect of the tubes on particle correlation is significant
only when the tubes stay close enough to the boundary of the system.
Therefore, instead of decomposing around Gaussian distribution,
we treat the IC naturally as superposition of peripheral high-energy tubes on top of the background.
To further simplify the problem, studies were carried out on the transverse hydrodynamic expansion
of a system consisting of only one tube located close to the surface of a smooth
background.\footnote{See Ref.\cite{sph-corr-7} for extension to the case with more than one tube.}
The resulting single-particle azimuthal distribution turned out to possess not a single peak as one might expect,
but two peaks symmetrically located on both sides of the original hot-spot position
(as shown in Fig.5 of ref.\cite{sph-corr-3})\footnote{A similar idea has been studied in Ref.\cite{ph-corr-shuryak-1}, but since the authors treated the tube as a small perturbation, the flow deflection may not be obtained, and this might explain why their awayside structure in two-particle correlation is very broad as compared with data.}.
This can be clearly seen as due to the deflection of the collective flow generated by the background into two directions,
caused by the explosion of the high-energy peripheral tube.
We therefore refer to the resulting hydrodynamic flow in response to high-energy tube as ``shadowing'' effect of the tube.
It can be shown straightforwardly that this causally connected double peaked distribution
eventually gives rise simultaneously to the desired ``ridge'' and ``shoulder'' structures (see Figs.$\,$6 and 7 of ref.\cite{sph-corr-3}).
Comparing to the triangularity-triangular flow interpretation,
the model does not require a one-particle distribution with three peaks to form the ``shoulder''.
Since the flow is deflected locally by the tube, only two peaks are formed.
In our picture, the hadrons giving rise both to the ``ridge" and ``shoulder'' structures are causally connected,
since they all come from the collective flow which was deflected by the very same tube.
In this context, peripheral-tube model considers ``ridges'' as a local effect, the generated flow is causally related.
Below, we will discuss more consequences of this description.

The results in section III show that linear hydrodynamic response partially breaks
and the coupling between cumulants of different orders can not be ignored.
Both of them cause the production of extra flow harmonics.
The deviation from linearity manifests itself as flow fluctuations when one performs event-by-event simulations.
These results are not at all surprising if one considers that hydrodynamics is a non-linear theory in nature.

In what follows, we show that above results may be intuitively understood in terms of the peripheral-tube model.
To do this, a particular set of IC is constructed in such a way that
it does not possess sextupole component but generates sizable triangular flow through hydrodynamic evolution.\footnote{A preliminary discussion has been presented by F.Gardim in ISMD2011 meeting\cite{sph-vn-2}.}
It is achieved based on the following considerations.
We have shown that the magnitude of correlation is a function of the position of the high-energy tube\cite{sph-corr-4}.
It decreases as the distance between the tube and the center of the background, $r$, decreases
and becomes negligible when $r<4.0$ fm in 200$\,A$GeV Au+Au collisions.
Therefore high-energy tubes placed closer to the center will contribute very little to the collective flow,
meanwhile they may contribute to the initial eccentricities according to the definition, Eq.(\ref{dcumulants}).
The energy distribution of the IC is shown in Fig.10.
One places a high-energy tube close to the boundary so that it affects the flow in the same way as it does in peripheral-tube model,
while another three tubes were placed closer to the axis of the background to give little contribution to the flow.
The angular positions and heights of the three inner tubes are adjusted so that the resultant initial sextupole is zero, $\varepsilon_3 = 0$.
By starting from these IC with zero $\varepsilon_3$ and computing the hydrodynamic expansion,
we correctly obtain the two-particle correlation with ``ridge'' and ``shoulder'' structures, as observed experimentally.
In Fig.11 we plot the corresponding initial eccentricities as well as the resulting flow components.
One sees that triangular flow in this case is quite big, and the flow harmonics look similar to those in Fig.7.
Though the above example is specially devised, and can not be treated as a general case,
it helps to illustrate the characteristics of peripheral-tube model
and it is particularly useful to give descriptions in those cases when linearity breaks down.
In such an approach, instead of decomposing collective flow into harmonic coefficients
and study their possible linear dependence on eccentricities,
we view the effect of fluctuations in terms of those of individual peripheral hot spots.
The physical outcome is therefore studied in terms of parameterization of the model,
i.e., height, size and total number of the tubes etc.

Before closing this section, we would like to give a few comments
on the causal connection of the flow and the correlations between event planes in the peripheral-tube model.
In addition, some possible implications on the difference from triangularity-triangular flow description will be discussed.
It has been shown by Glauber Monte Carlo models
that odd order event planes are largely not correlated to the reaction plane for non-central collisions\cite{glauber-en-3,hydro-eve-3,hydro-v3-4}.
In the peripheral-tube model, the situation is more delicate.
Since the part of the collective flow generated locally by tube is causally related,
the corresponding event planes, inclusively that of the triangular flow, are automatically correlated to the location of the tube.
However, one must bear in mind that the flow due to tube only accounts for a small proportion of the overall flow of the whole system.
Straightforward calculation shows that these correlations (between event planes) do not easily present themselves in di-hadron correlation.
Nevertheless, they might be measured directly or by three particle or 2+1 particle correlations.
It was shown that event plane correlation, which is obtained naturally in peripheral-tube model,
can be used to explain the trigger-angle dependence of di-hadron correlation \cite{sph-corr-ev-2,sph-corr-ev-4}.
In fact, efforts have been made in the study of connection between participant planes \cite{glauber-en-3,glauber-en-1,glauber-en-2},
in terms of random geometrical fluctuations in Monte Carlo Glauber IC and linearity between eccentricities and flow harmonics.
To us, it is therefore legitimate to ask, to what extent does each mechanism account for the observed data.
This idea will be further explored in more details in a further work.

Nonlinearity have also been studied by other authors\cite{ph-vn-7,ph-vn-8,hydro-vn-1,hydro-vn-2,sph-vn-3,hydro-vn-3}.
Though from a different approach, our results are consistent with those obtained by others.
A novel feature of the present study lies in the fact that we have focused only some specific cumulant, and ask how it may affect the resulting flows.
For instance, we investigated how $\epsilon_1$ alone will produce flow components besides $v_1$.
In terms of cumulant expansion\cite{hydro-vn-2}, the appearance of higher order flow harmonics (such as $v_2$, $v_3$ in this case) 
can be attributed to the $\epsilon_1$. 
And in their mode, it is an $n$-th order effect.
On the other hand, the production of $v_3$ from IC without any $e_3$ can also be understood by cumulant expansion.
In our approach, we understand this result as a natural consequence of our first proposed test.
In terms of cumulants \cite{hydro-vn-2}, the produced $v_3$ can be attributed to the contributions from $\epsilon_1$ and $\epsilon_2$.
Our results is also consistent with the correlations between $\epsilon_n$ and $v_n$ studied in ref.\cite{hydro-vn-3}.
It was shown that the linear relation between $\epsilon_n$ and $v_n$ becomes spread out when $n$ increases.
Based on the above calculations, this can be understood quite natrually because when the order $n$ increases, 
there are more contributions from those lower order cumulants.
We also explored the effect of eccentricity plane rotations and used it to study the coupling between different cumulants,
this can be seen as a source for flow fluctuations besides those due to the fluctuation of the magnitudes of cumulants.
In the literature, such angular correlation was also explored using different approaches in terms of correlations 
between participant planes \cite{glauber-en-1,glauber-en-2} or event plane \cite{hydro-phin-1}.
However, throughout the study, we did not presume any linearity between eccentricities and momentum anisotropies,
neither of produced hadron, nor of fluid itself.
Consequently, we did not attempt to determine any predefined response coefficient.
Though in this work, we have not considered viscosity, 
but the method employed in the current study can be easily generalized to the cases with the presence of viscosity.

In summary, we carried out a systematic study on the connection between cumulants
and the flow harmonics using the hydrodynamic code NeXSPheRIO.
In order to study the hydrodynamic response to the inicial eccentricity more transparently, we conducted three types of calculations.
Precisely, the hydrodynamic evolutions of initial conditions composed of individual cumulants are studied.
Though in general, linearity is approximately obtained, it is found that, in some cases, small extra flow harmonics of higher orders are produced,
deviating from the proposed linearity between eccentricities and flow coefficients.
In particular, we have found that $e_1$ alone may produce higher order harmonics,
and IC without sextupole component still may generate sizable triangular flow, and two-particle correlation as observed experimentally.
General conclusion is further drawn from quantitative event-by-event analysis.
It is argued that above results can be intuitively understood by devising a toy IC motivated by the peripheral-tube model.

\section{V. Acknowledgments}
We thank for valuable discussions with Klaus Werner, Paul Sorensen and Jiangyong Jia.
We acknowledge funding from Funda\c{c}\~ao de Amparo \`a Pesquisa do Estado de S\~ao Paulo, 
FAPESP, Funda\c{c}\~ao de Amparo \`a Pesquisa do Estado de Minas Gerais, FAPEMIG,
and Conselho Nacional de Desenvolvimento Cient\'{\i}fico e Tecnol\'ogico, CNPq.

\bibliographystyle{h-physrev}

\bibliography{references_qian}{}

\setcounter{figure}{0}

\begin{figure}[!htb]
\vspace*{-0.0cm}
\includegraphics[width=350pt]{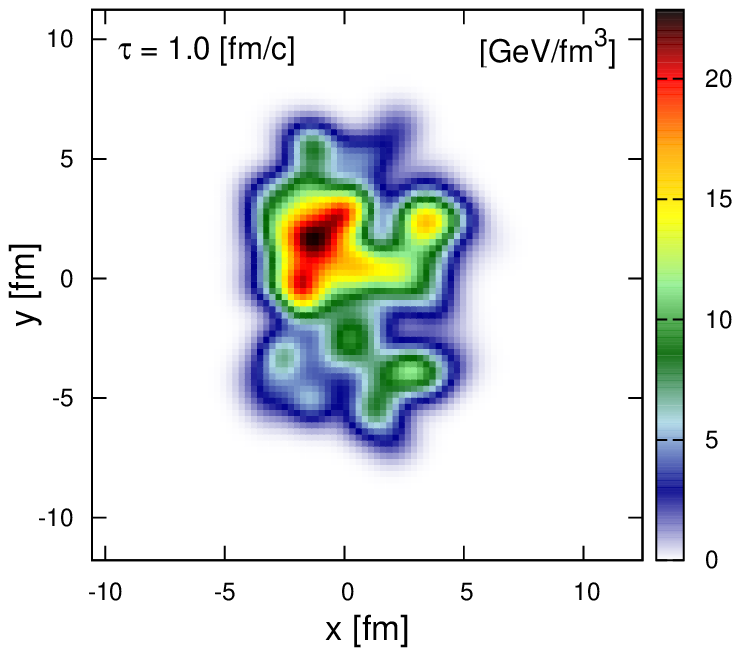}
\vspace*{-0.0cm}
\caption{(Color online) Energy distribution of one random NeXuS event.}
\label{fig1}
\end{figure}

\begin{figure}
\begin{tabular}{cc}
\begin{minipage}{200pt}
\includegraphics[width=220pt]{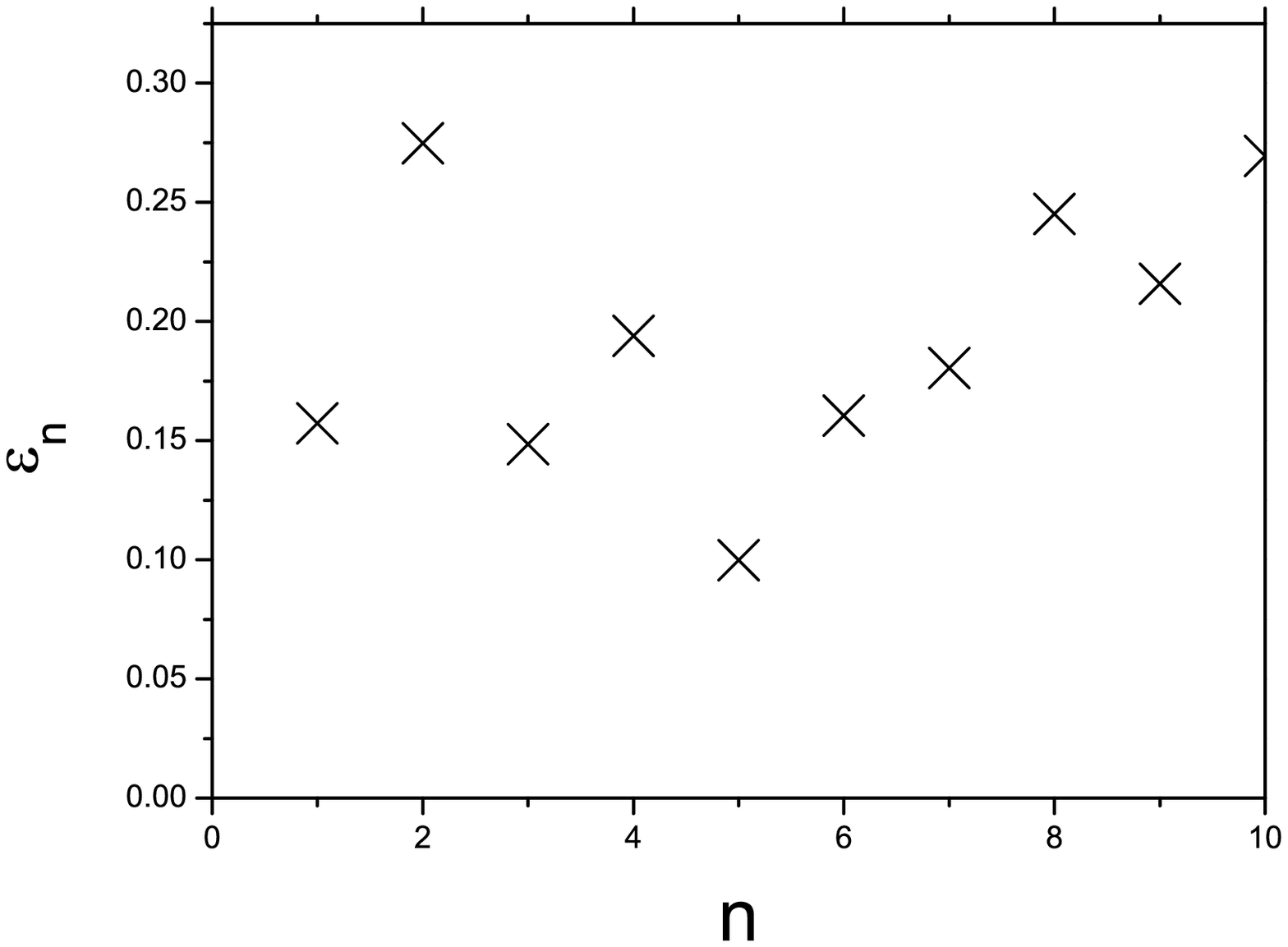}
\end{minipage}
&
\begin{minipage}{200pt}
\includegraphics[width=220pt]{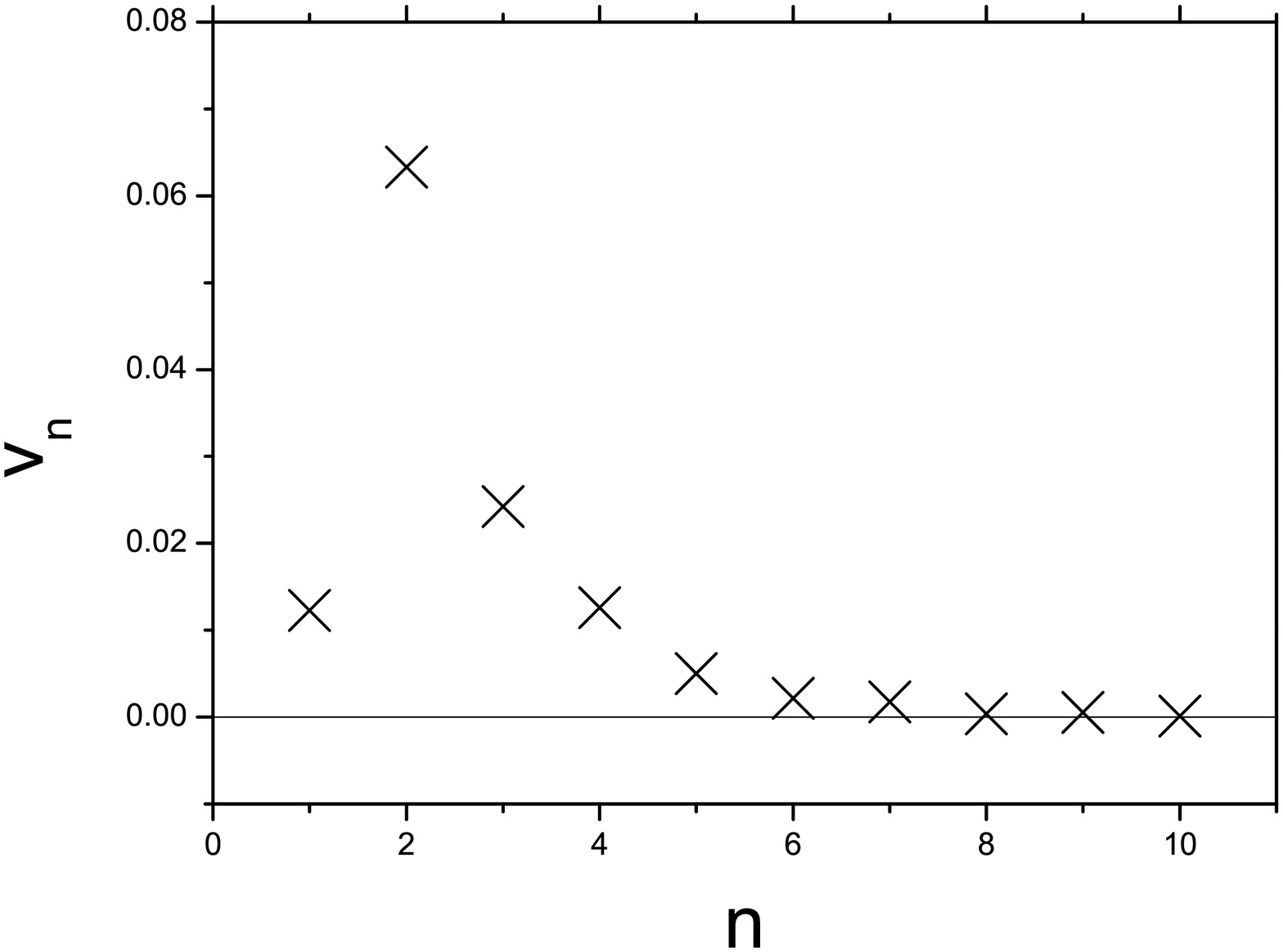}
\end{minipage}
\end{tabular}
\caption{Eccentricities $\varepsilon_n$ and flow harmonics $v_n$ of the same NeXuS event shown in Fig.1 as a function of $n$.}
\label{fig2}
\end{figure}

\begin{figure}
\begin{tabular}{cc}
\begin{minipage}{200pt}
\centerline{\includegraphics[width=300pt]{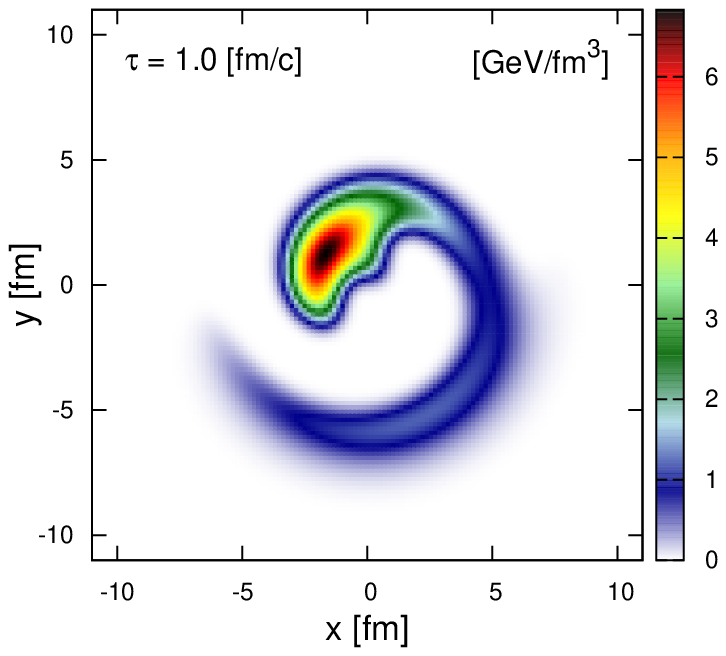}}
\end{minipage}
&
\begin{minipage}{200pt}
\centerline{\includegraphics[width=300pt]{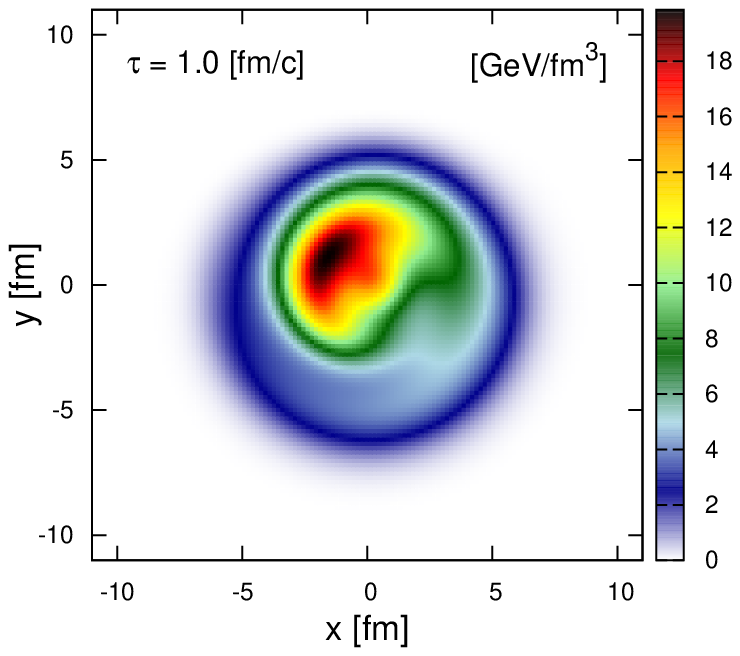}}
\end{minipage}
\\
\begin{minipage}{200pt}
\centerline{\includegraphics[width=300pt]{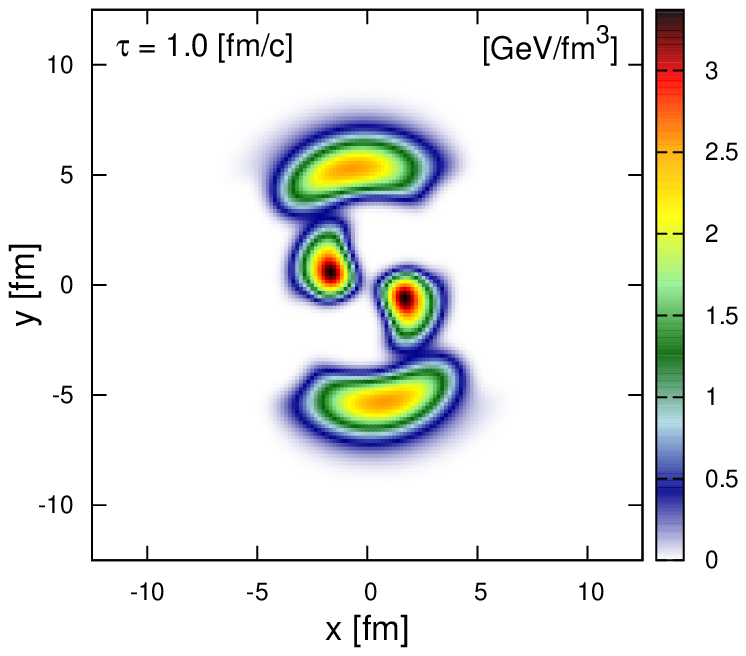}}
\end{minipage}
&
\begin{minipage}{200pt}
\centerline{\includegraphics[width=300pt]{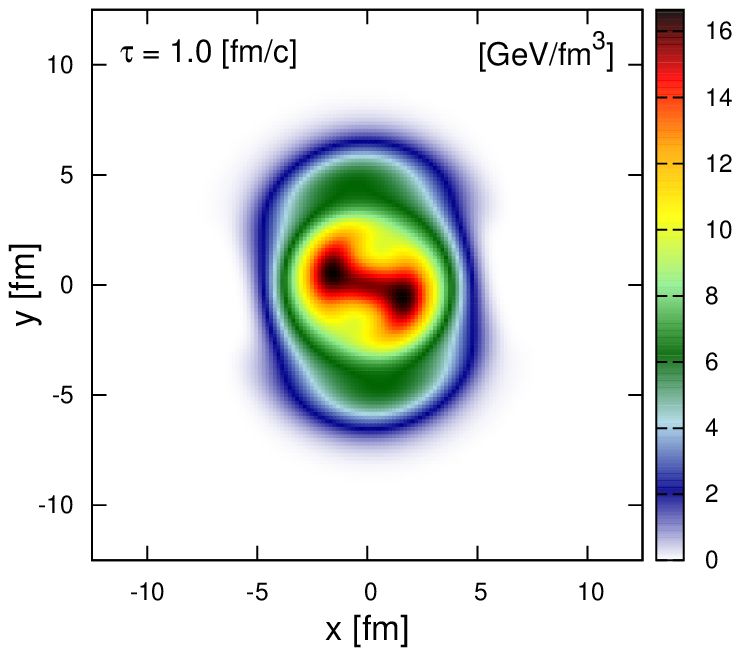}}
\end{minipage}
\\
\begin{minipage}{200pt}
\centerline{\includegraphics[width=300pt]{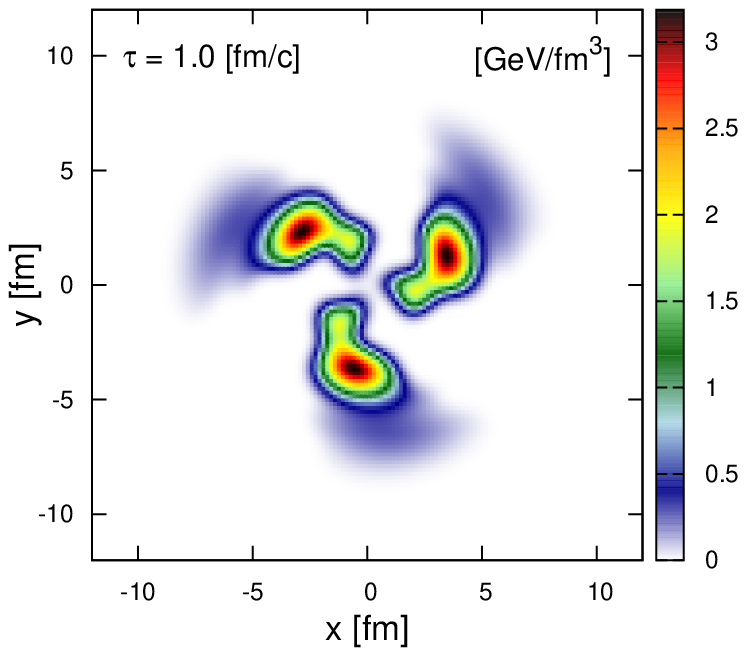}}
\end{minipage}
&
\begin{minipage}{200pt}
\centerline{\includegraphics[width=300pt]{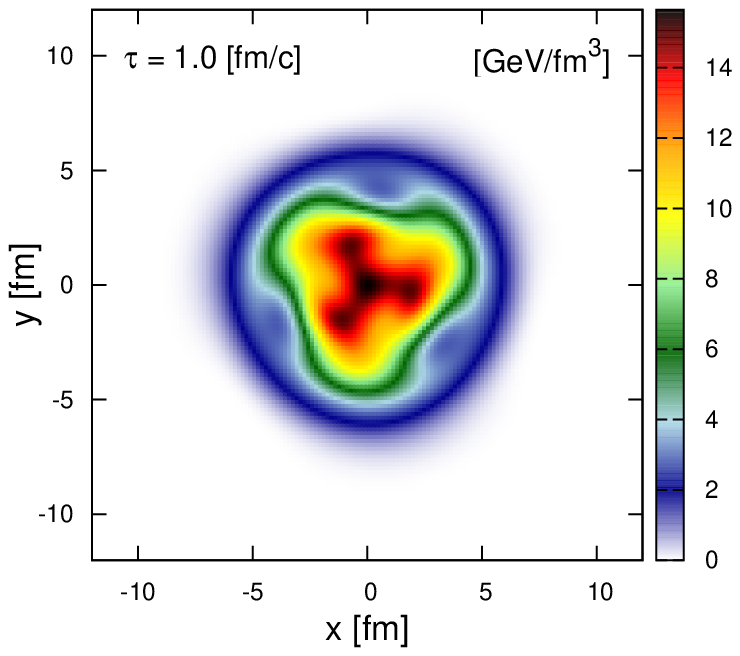}}
\end{minipage}
\\
\end{tabular}
 \caption{(Color online) Energy distributions $e_n(r,\phi)$ of the three devised event from the first proposed test. On the l.h.s. from top to bottom, positive parts of energy distributions of $e_1$,
$e_2$ and $e_3$. On the r.h.s. from top to bottom, energy distributions of $e_0+e_1$,
$e_0+e_2$ and $e_0+e_3$.}
 \label{fig3}
\end{figure}

\begin{figure}
\begin{tabular}{cc}
\begin{minipage}{200pt}
\centerline{\includegraphics[width=220pt]{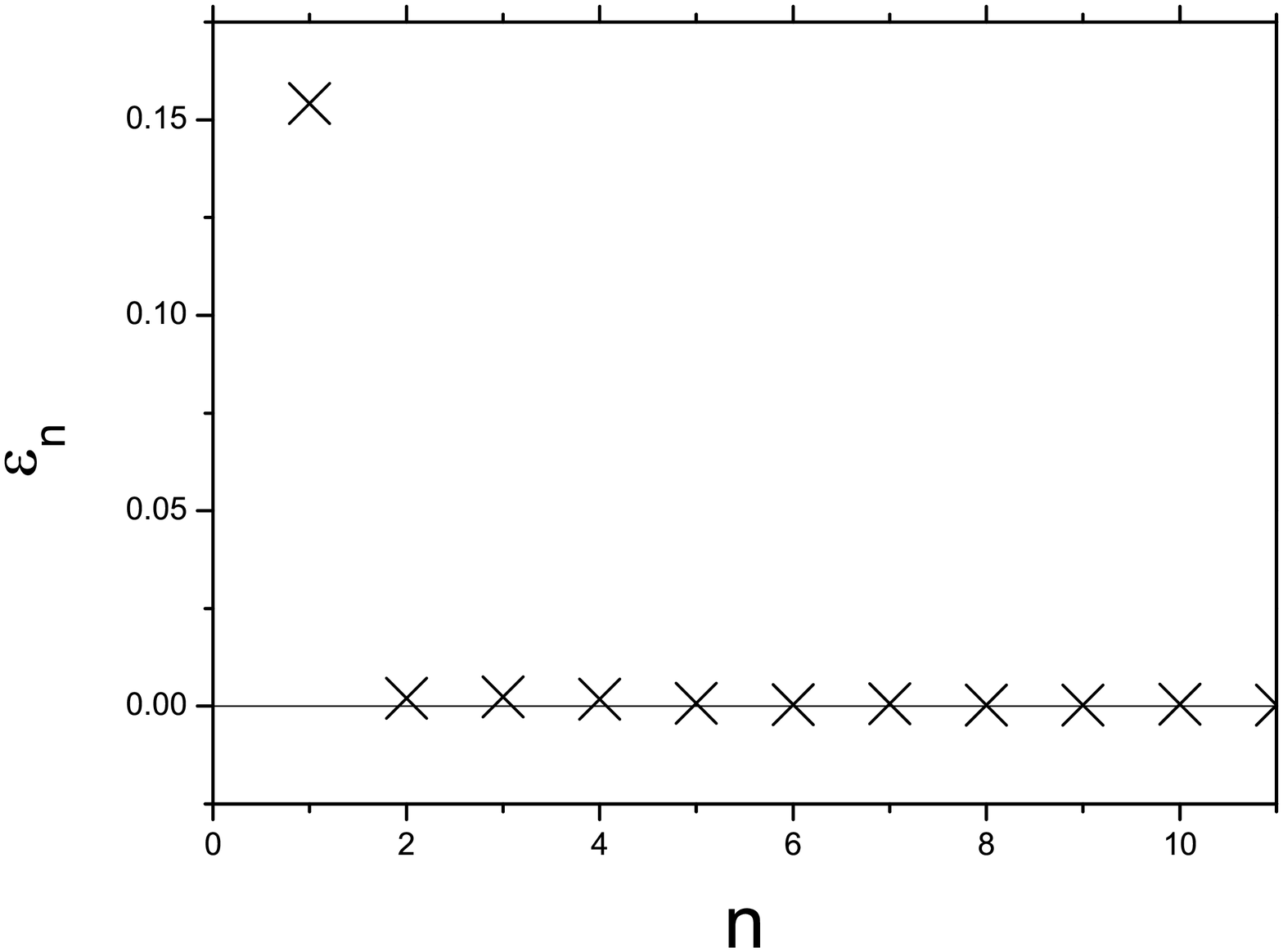}}
\end{minipage}
&
\begin{minipage}{200pt}
\centerline{\includegraphics[width=220pt]{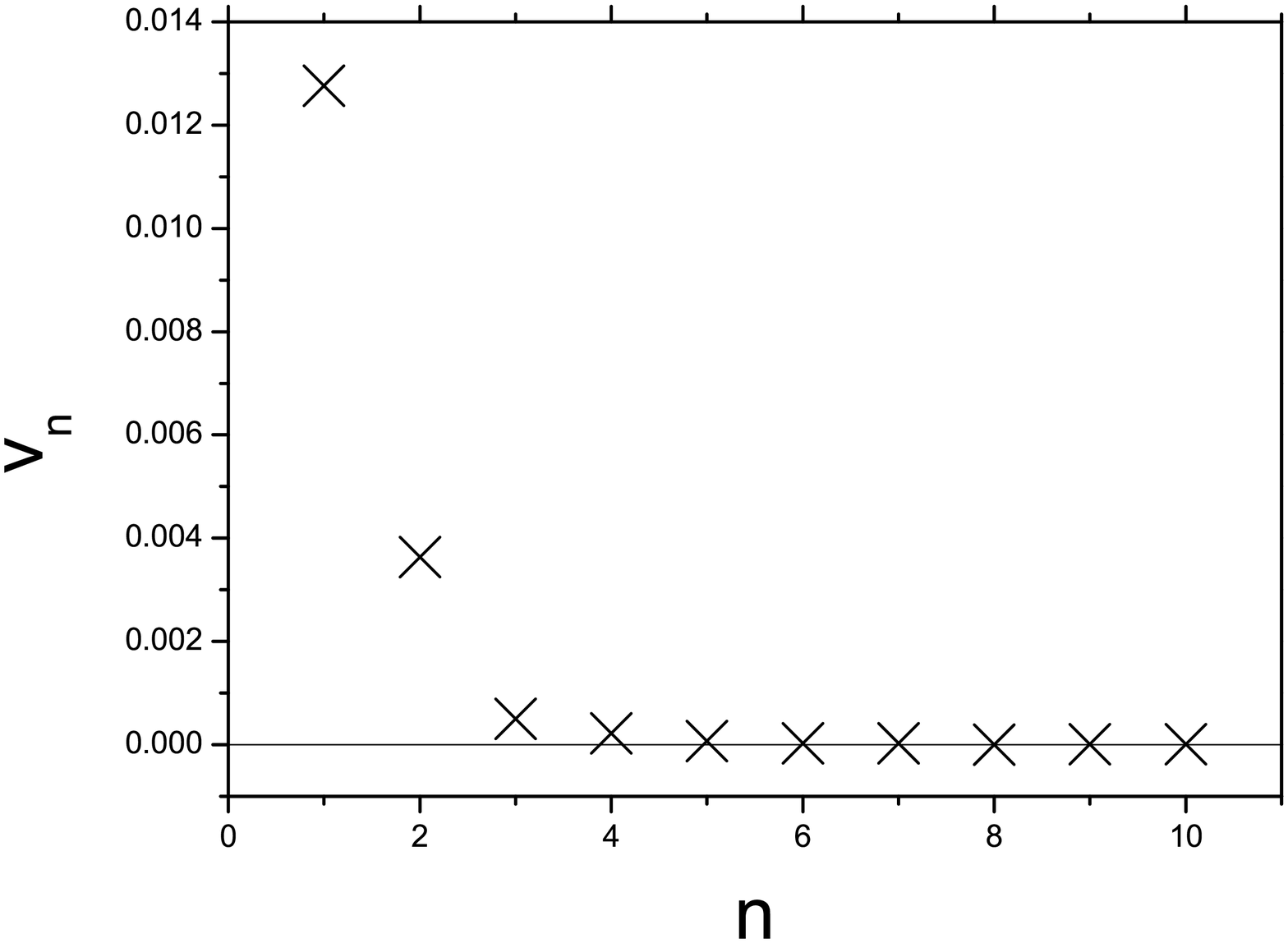}}
\end{minipage}
\\
\begin{minipage}{200pt}
\centerline{\includegraphics[width=220pt]{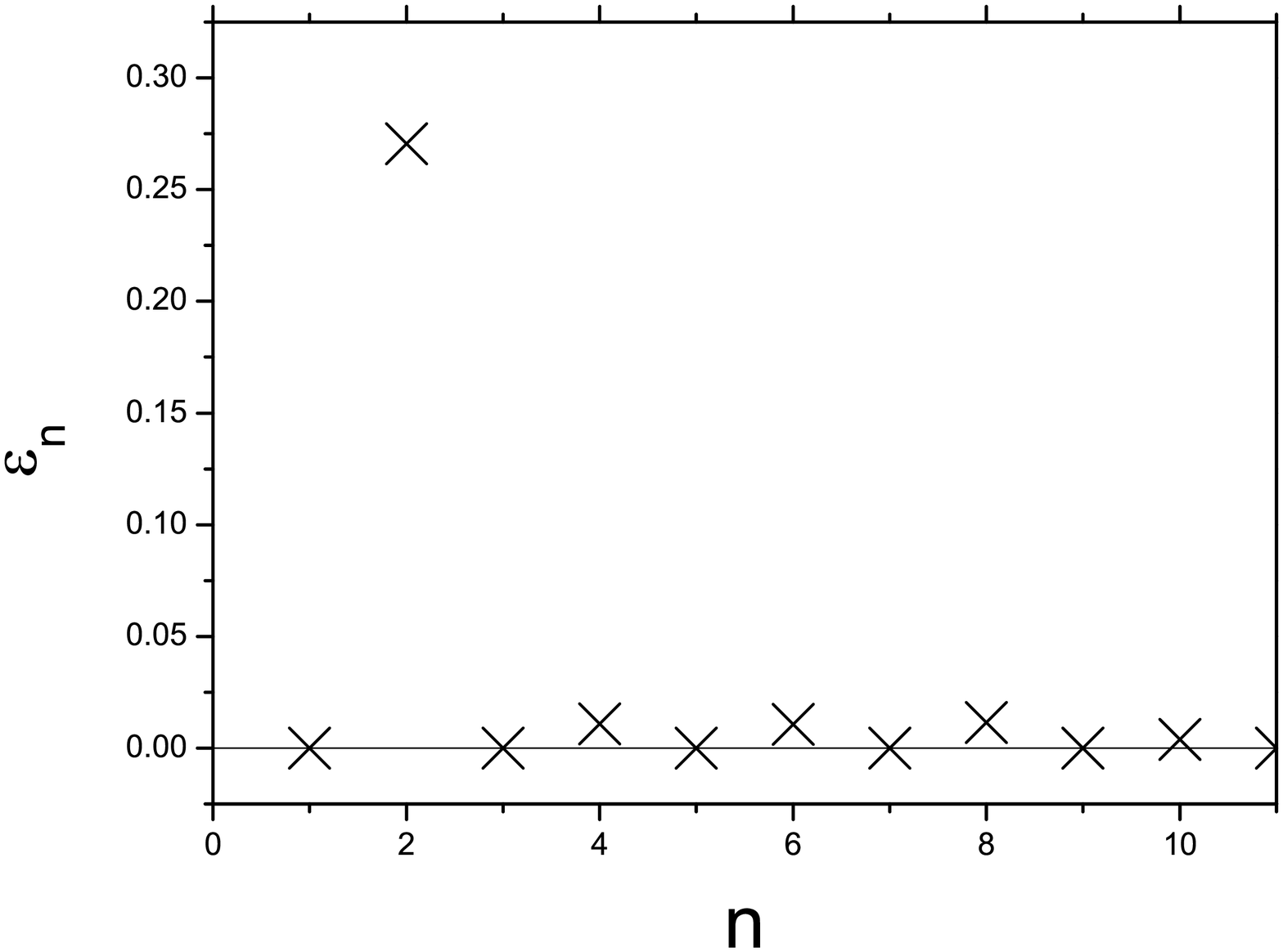}}
\end{minipage}
&
\begin{minipage}{200pt}
\centerline{\includegraphics[width=220pt]{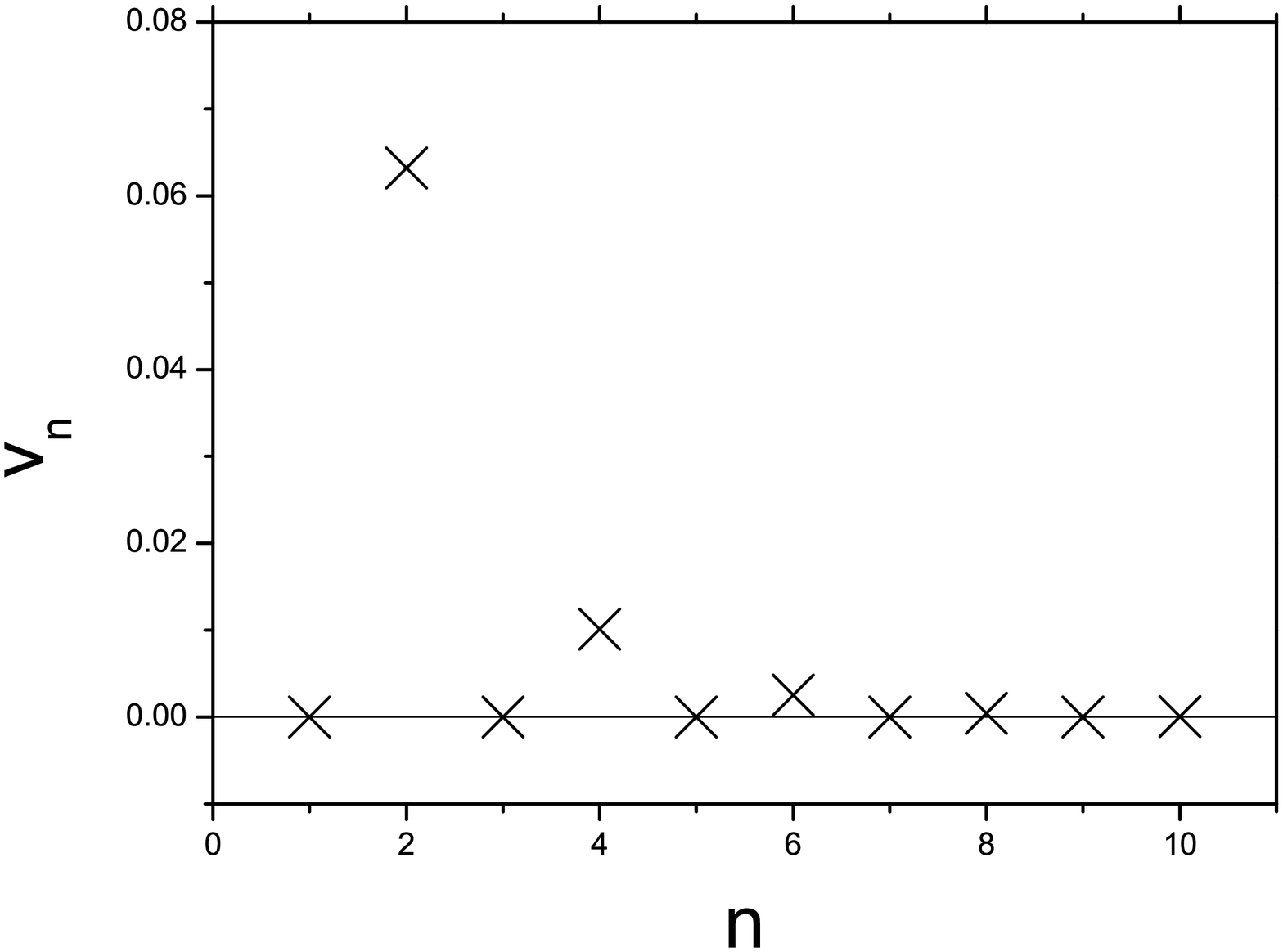}}
\end{minipage}
\\
\begin{minipage}{200pt}
\centerline{\includegraphics[width=220pt]{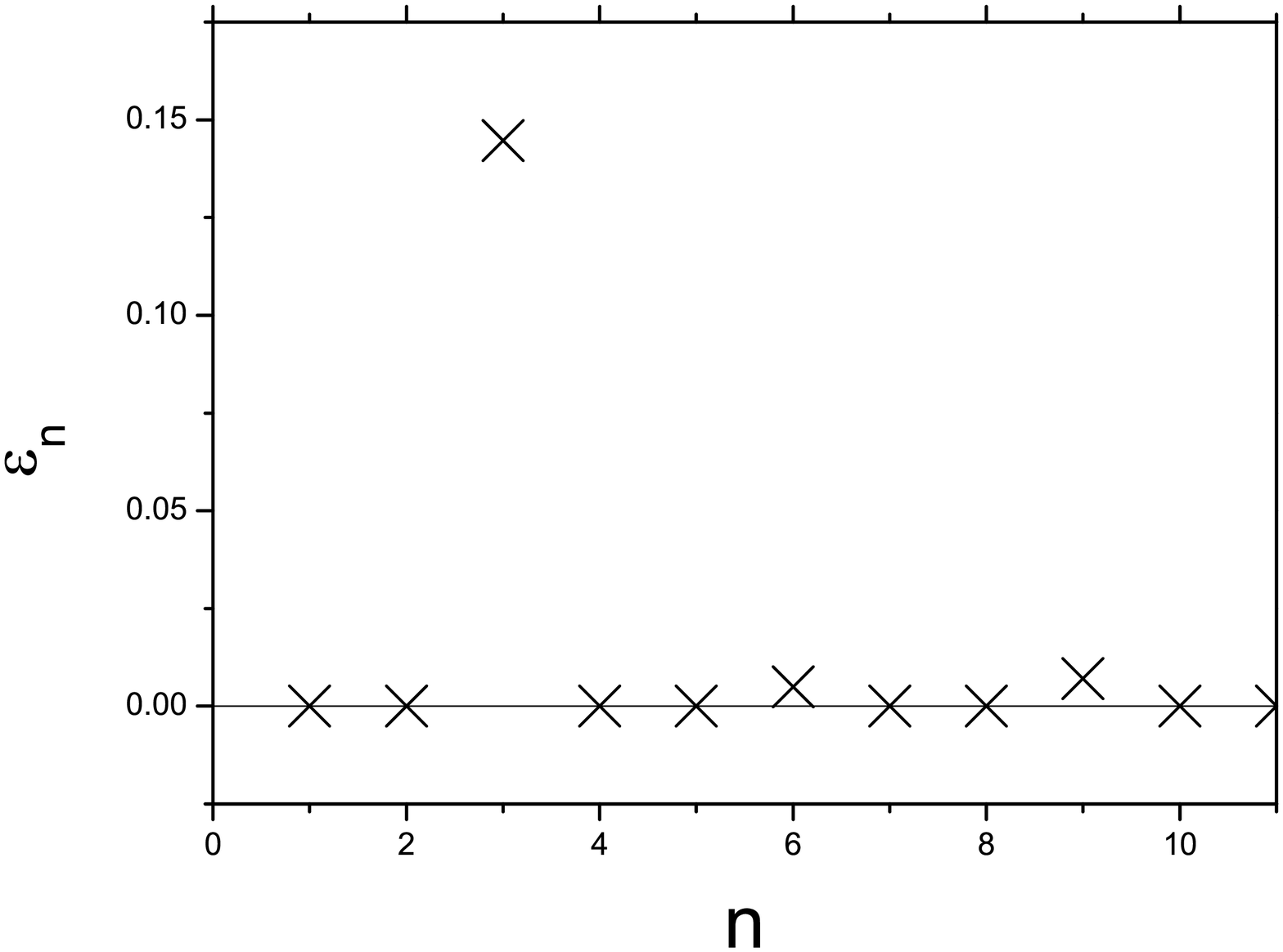}}
\end{minipage}
&
\begin{minipage}{200pt}
\centerline{\includegraphics[width=220pt]{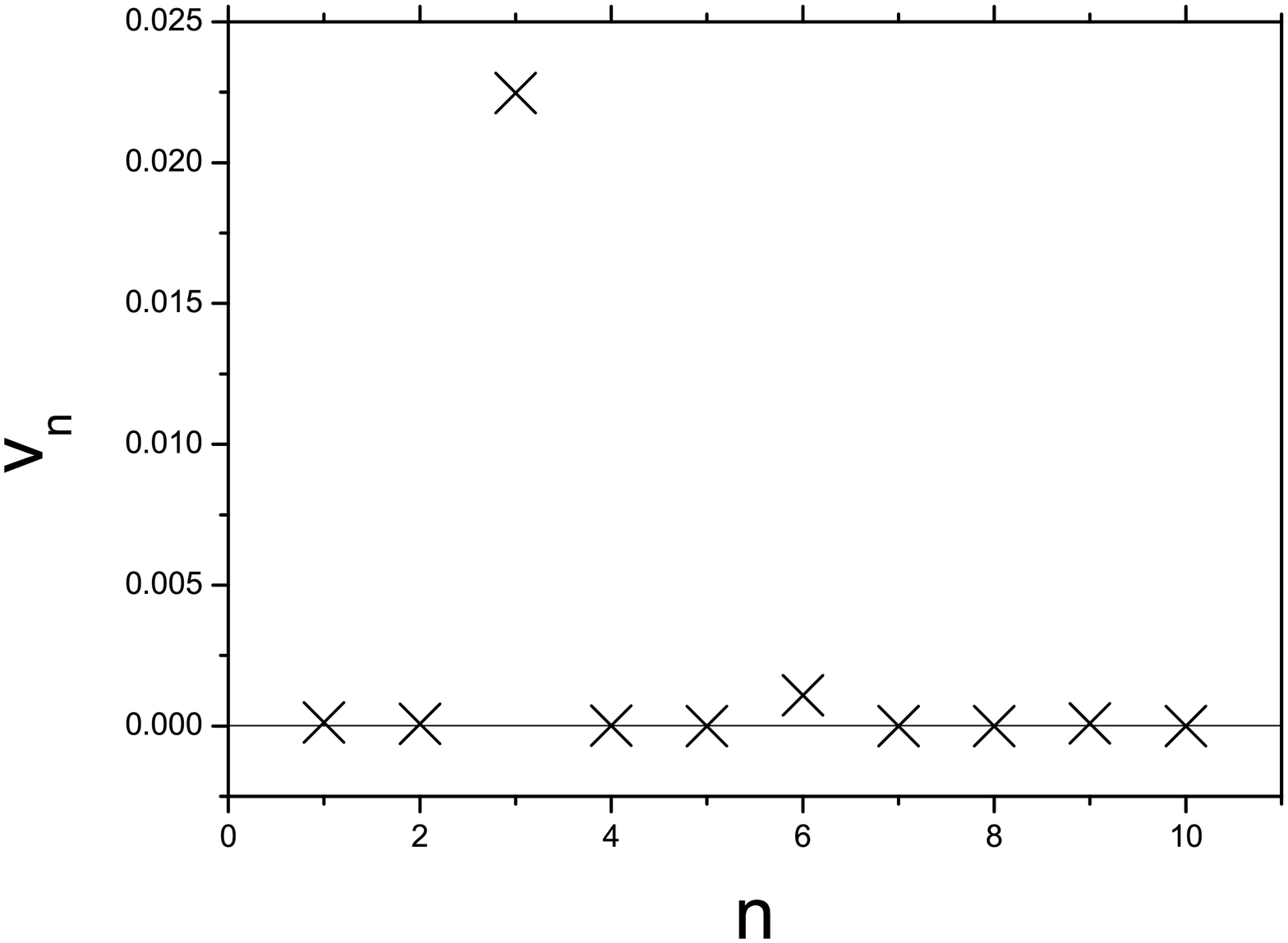}}
\end{minipage}
\\
\end{tabular}
 \caption{Initial eccentricities and the calculated flow harmonics of the three devised event. On the l.h.s. from top to bottom, cumulant components of $e_1$,
$e_2$ and $e_3$. On the r.h.s. from top to bottom, flow harmonics of $e_0+e_1$,
$e_0+e_2$ and $e_0+e_3$.}
 \label{fig4}
\end{figure}

\begin{figure}
\begin{tabular}{cc}
\begin{minipage}{200pt}
\centerline{\includegraphics[width=220pt]{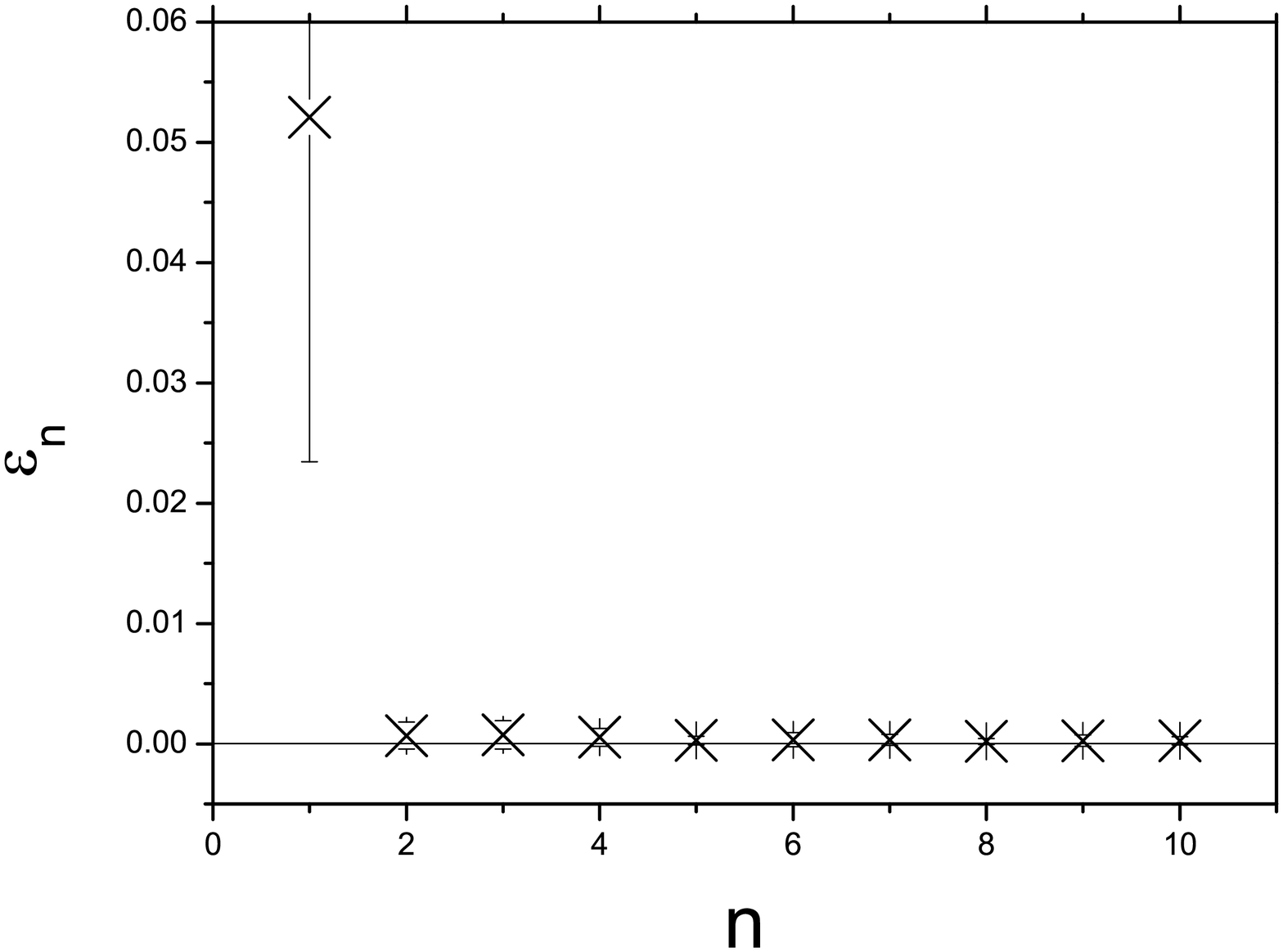}}
\end{minipage}
&
\begin{minipage}{200pt}
\centerline{\includegraphics[width=220pt]{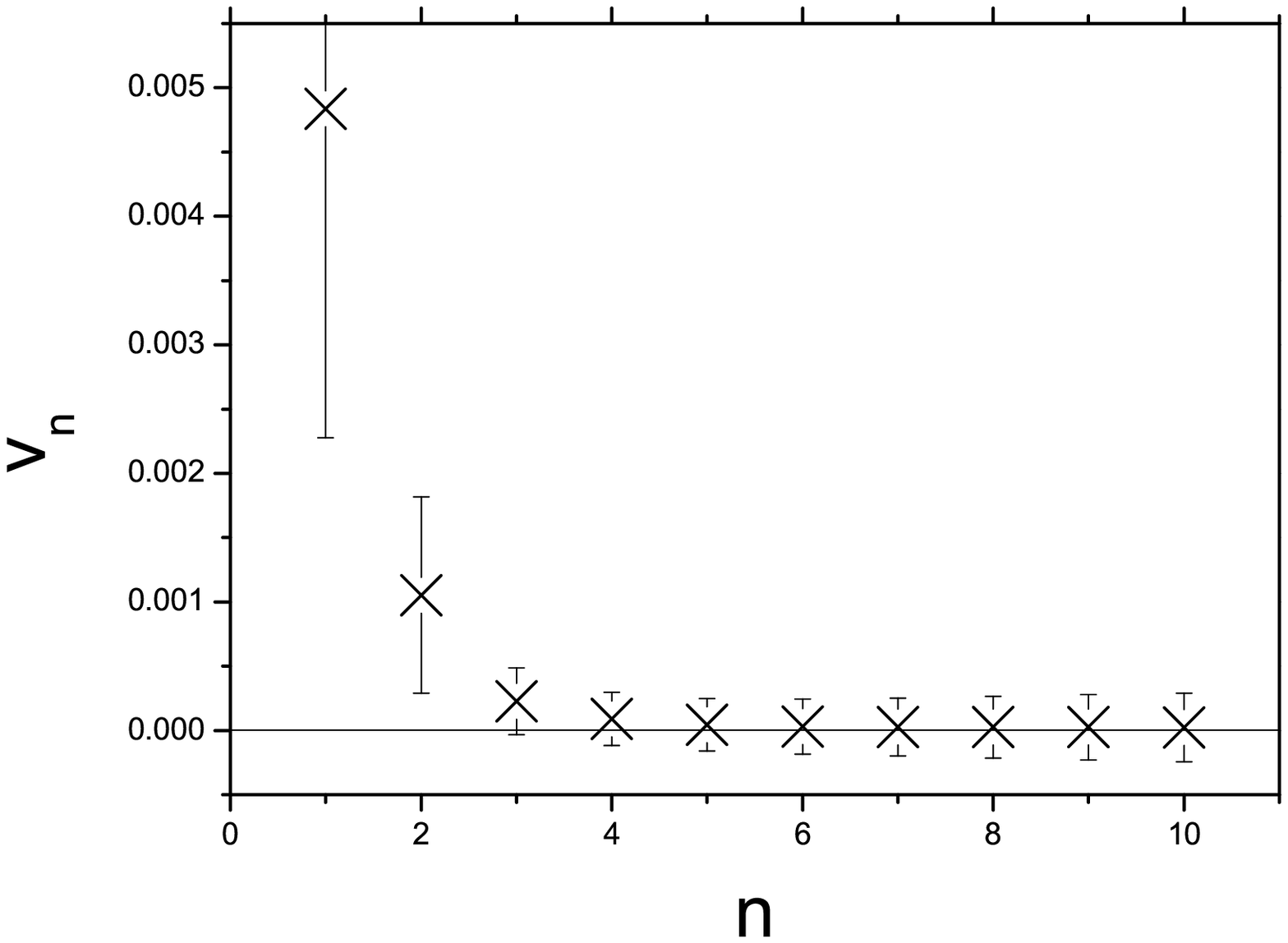}}
\end{minipage}
\\
\begin{minipage}{200pt}
\centerline{\includegraphics[width=220pt]{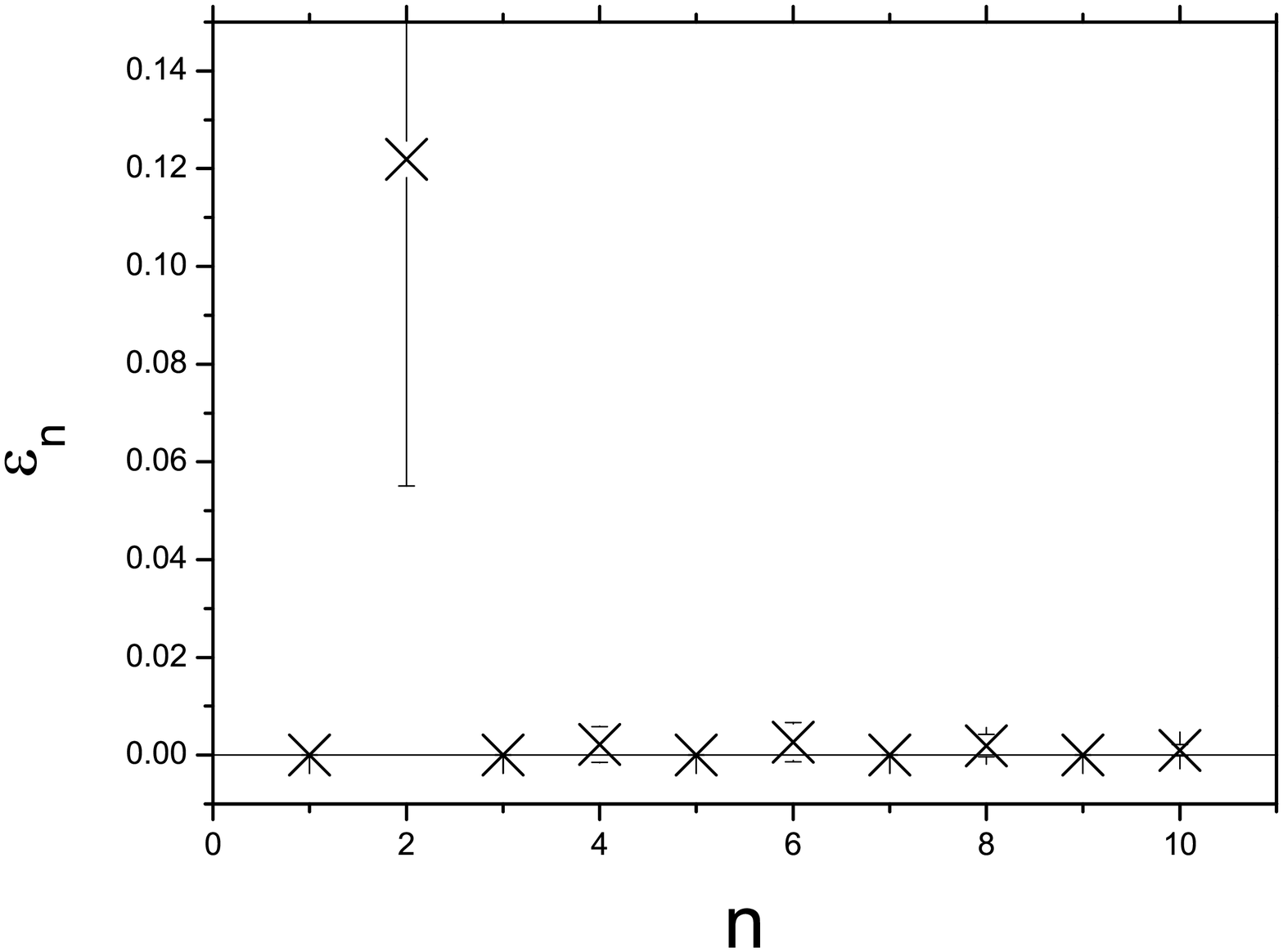}}
\end{minipage}
&
\begin{minipage}{200pt}
\centerline{\includegraphics[width=220pt]{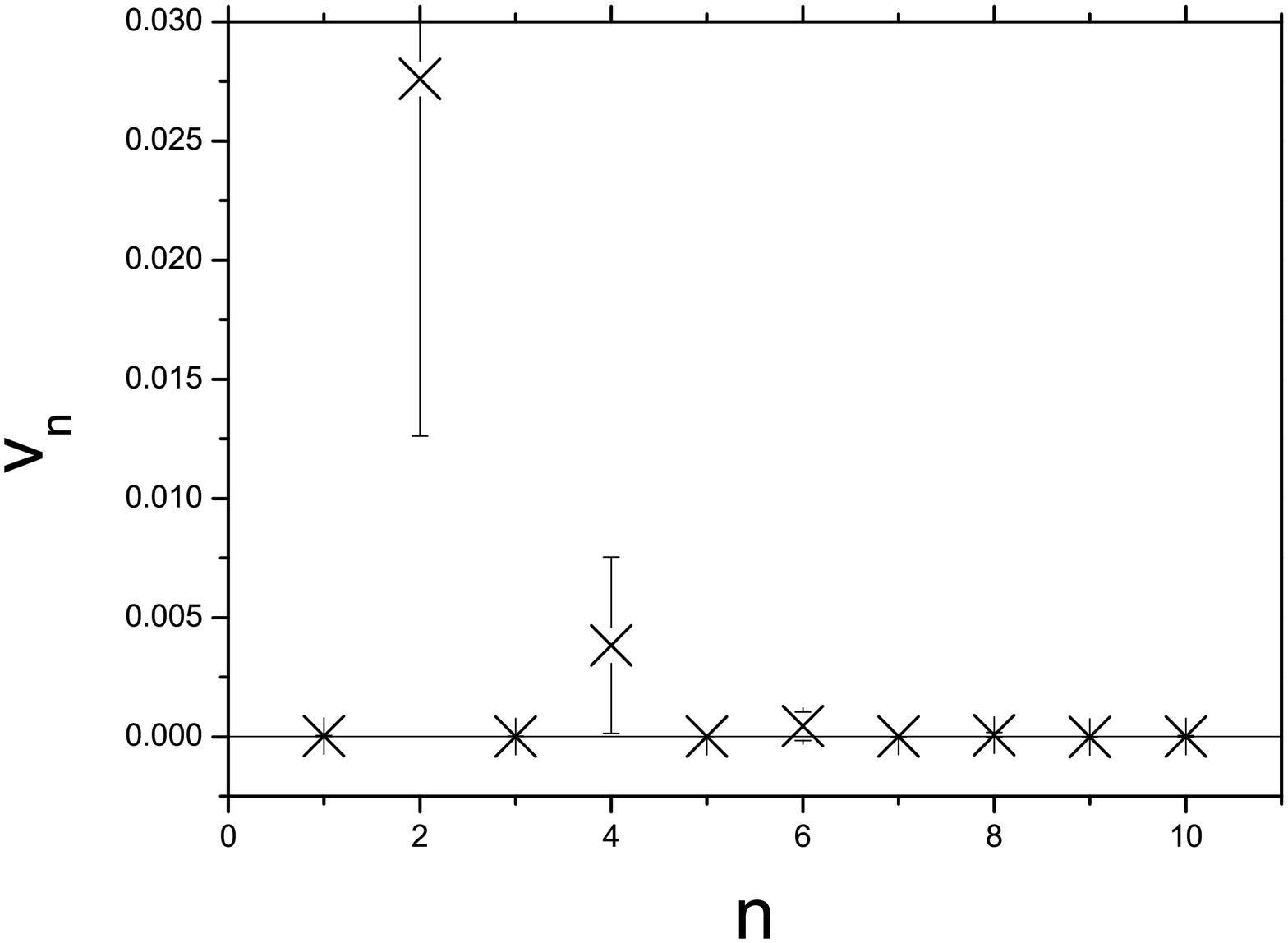}}
\end{minipage}
\\
\begin{minipage}{200pt}
\centerline{\includegraphics[width=220pt]{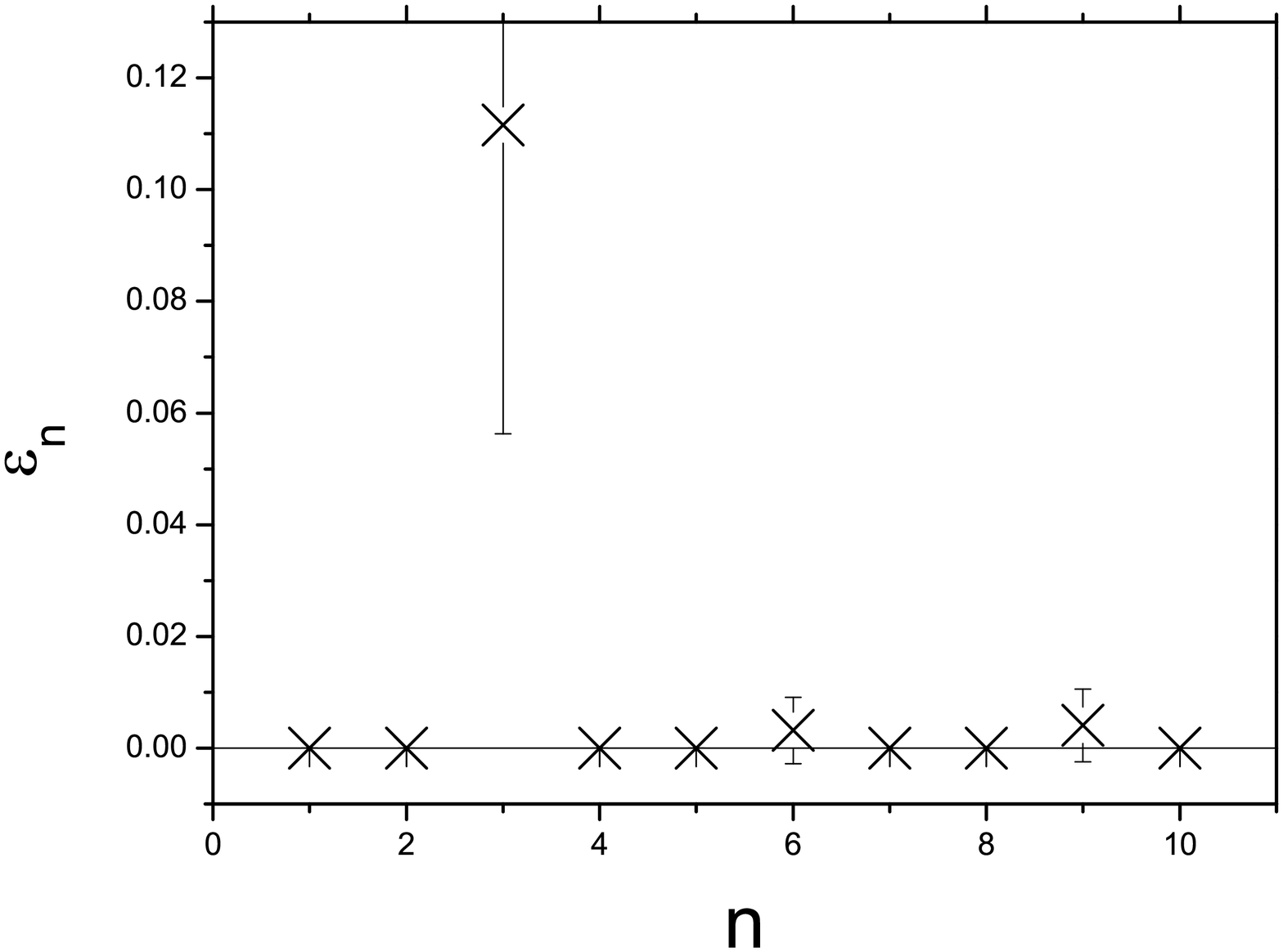}}
\end{minipage}
&
\begin{minipage}{200pt}
\centerline{\includegraphics[width=220pt]{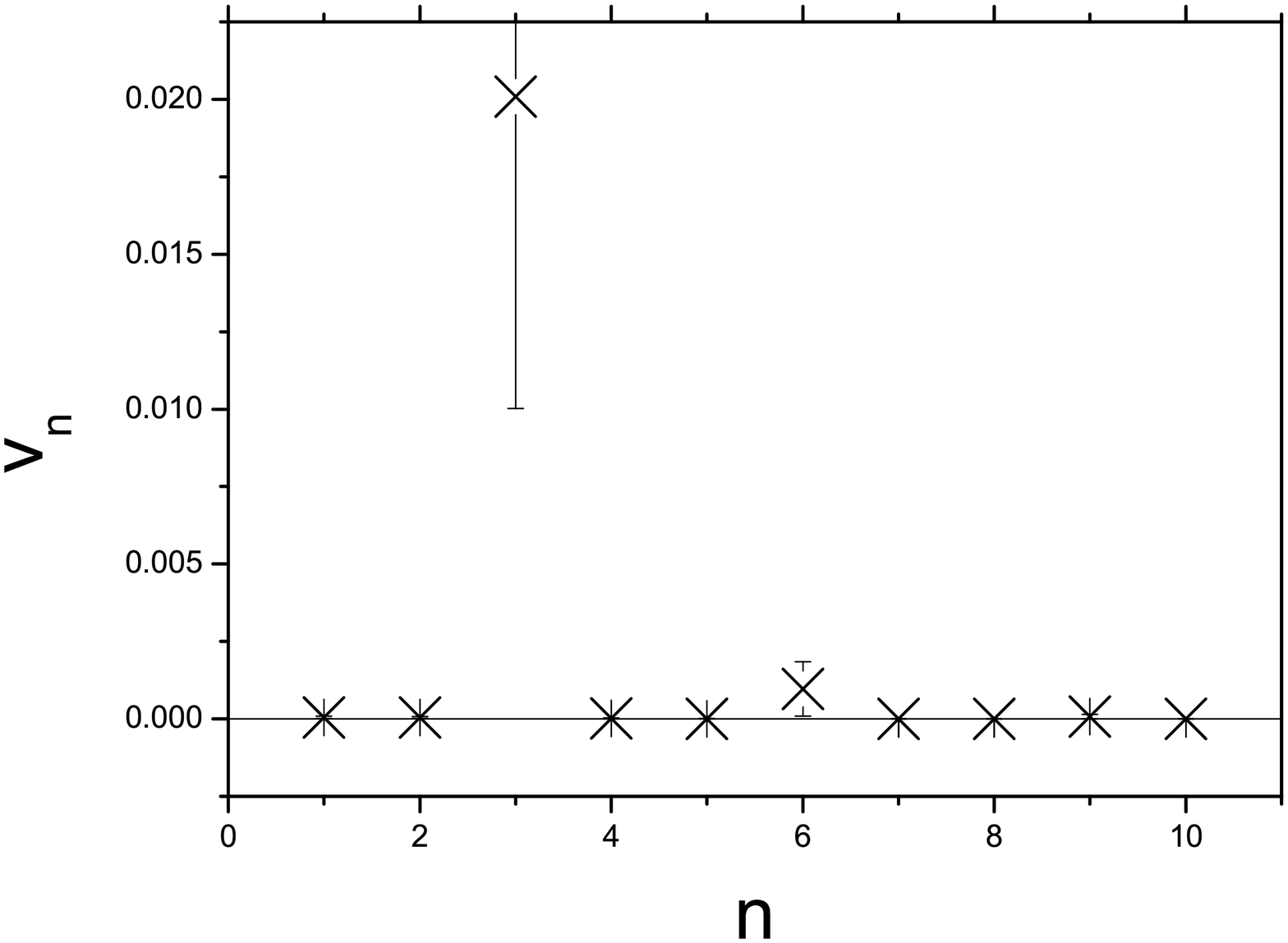}}
\end{minipage}
\\
\end{tabular}
 \caption{The same as Fig.4 but for results averaged over 150 random events, where error bars indicate the standard deviations.}
 \label{fig5}
\end{figure}

\begin{figure}[!htb]
\vspace*{-0.0cm}
\includegraphics[width=10.cm]{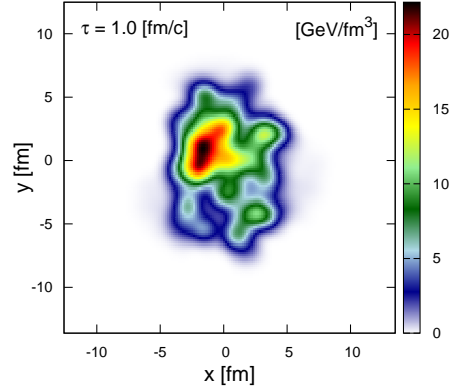}
\vspace*{-0.0cm}
\caption{(Color online) Energy distribution of the second proposed test, the event is generated by removing the $e_3$ from the original event shown in Fig.1.}
\label{fig6}
\end{figure}

\begin{figure}
\begin{tabular}{cc}
\begin{minipage}{200pt}
\centerline{\includegraphics[width=220pt]{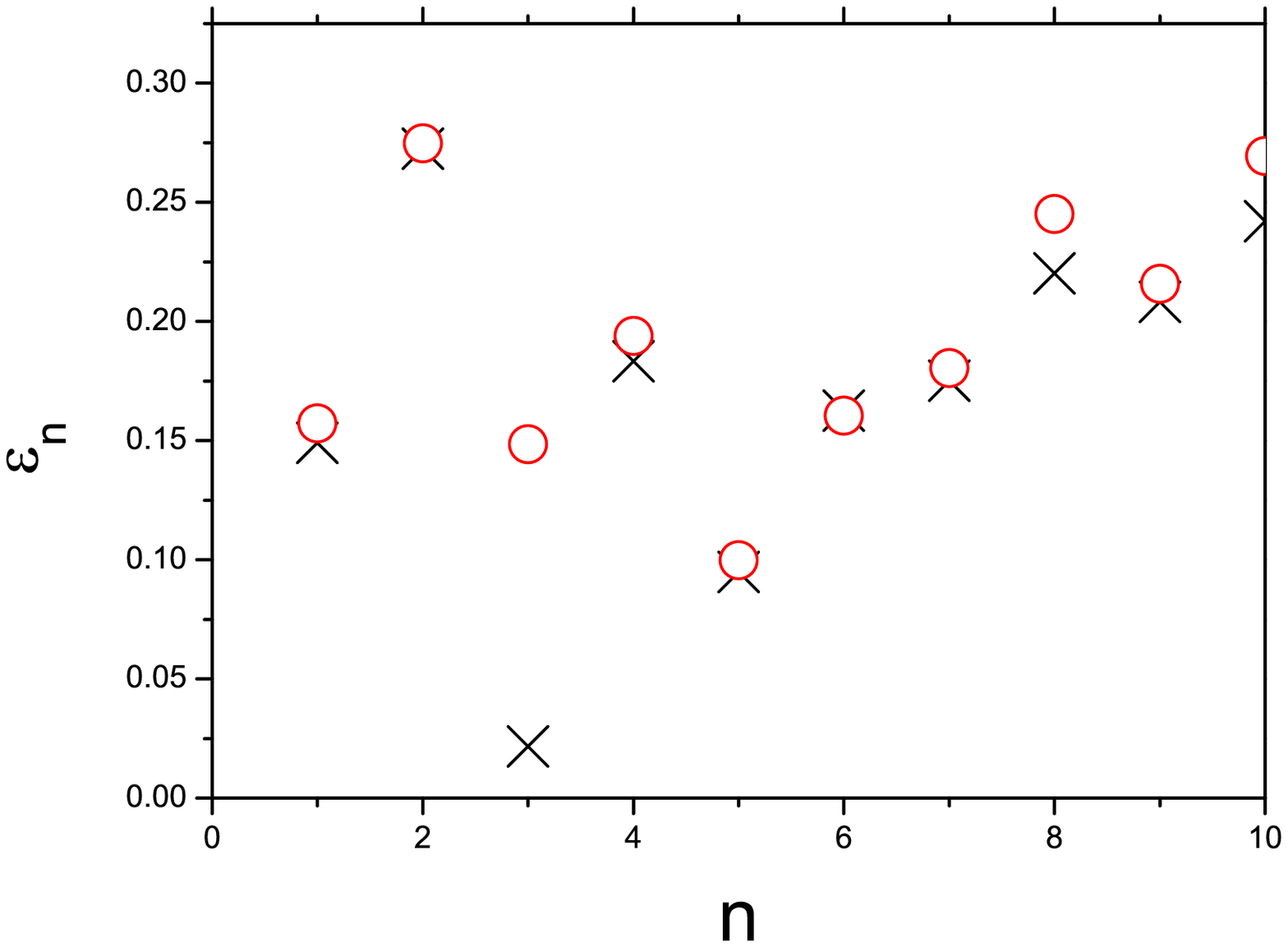}}
\end{minipage}
&
\begin{minipage}{200pt}
\centerline{\includegraphics[width=220pt]{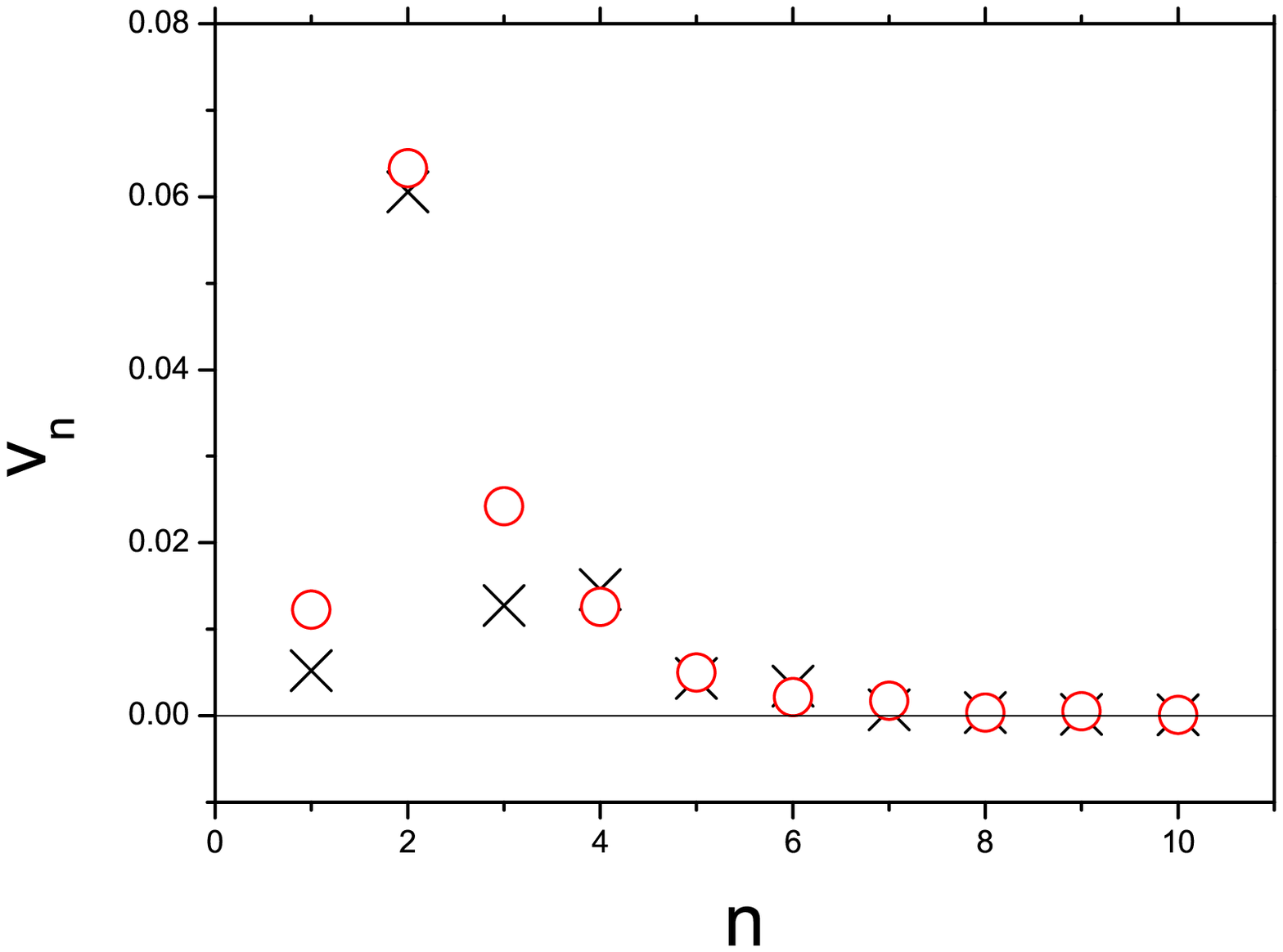}}
\end{minipage}
\\
\end{tabular}
\begin{minipage}{200pt}
\centerline{\includegraphics[width=220pt]{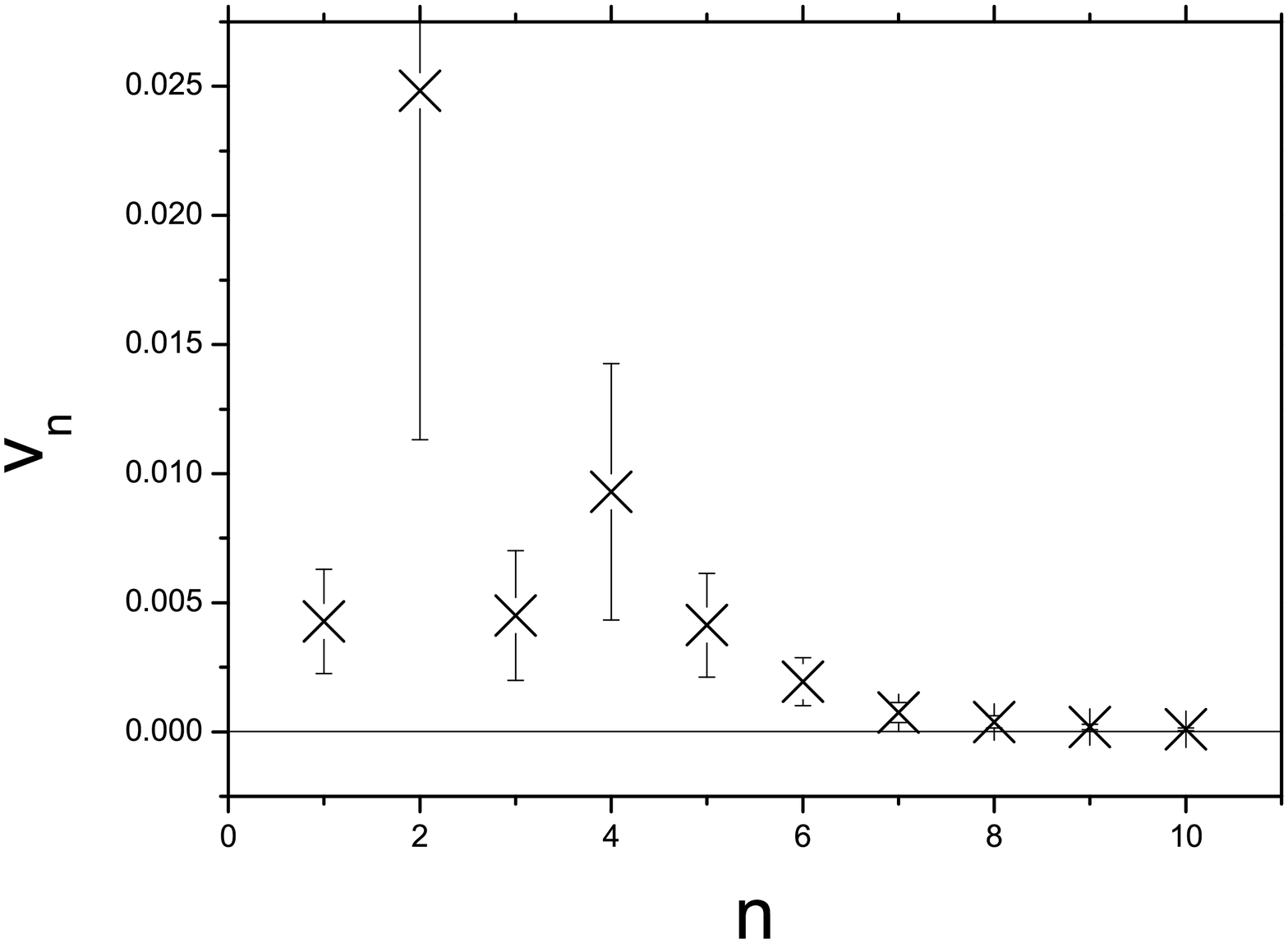}}
\end{minipage}
 \caption{(Color online) Eccentricities and calculated flow of the second proposed test, the black crosses are the calculated results of the second test and the red circles are those of the original event. The top left plot: initial eccentricities, the top right one: calculated flow harmonics, and the bottom one: calculated average flow harmonics from 150 events.}
 \label{fig7}
\end{figure}

\begin{figure}
\begin{tabular}{cc}
\begin{minipage}{200pt}
\centerline{\includegraphics[width=300pt]{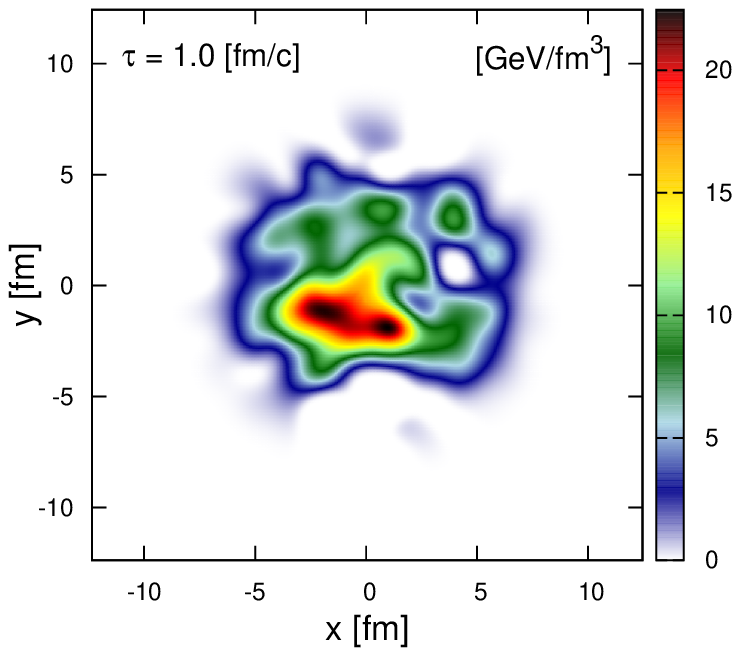}}
\end{minipage}
&
\begin{minipage}{200pt}
\centerline{\includegraphics[width=300pt]{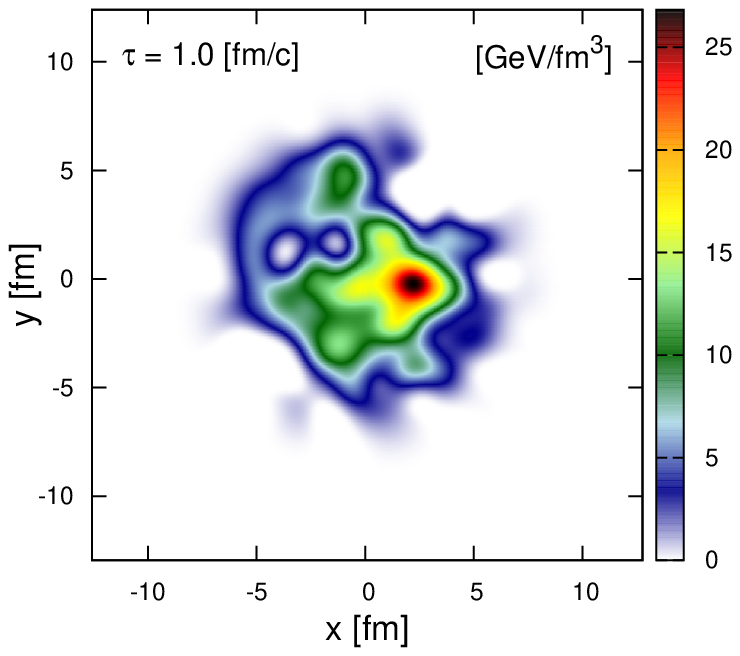}}
\end{minipage}
\\
\end{tabular}
 \caption{(Color online) Energy distribution of the third proposed test, these two events are obtained by randomizing the eccentricity planes of the original event shown in Fig.1.}
 \label{fig8}
\end{figure}

\begin{figure}[!htb]
\vspace*{-0.0cm}
\includegraphics[width=10.cm]{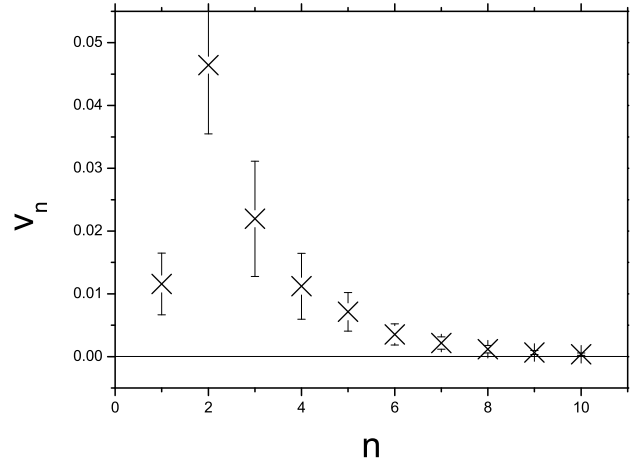}
\vspace*{-0.0cm}
\caption{Calculated flow harmonics of the third proposed test, the results are obtained by averaging 150 events and the error bars are the standard deviations.}
\label{fig9}
\end{figure}

\begin{figure}[!htb]
\vspace*{-0.0cm}
\includegraphics[width=10.cm]{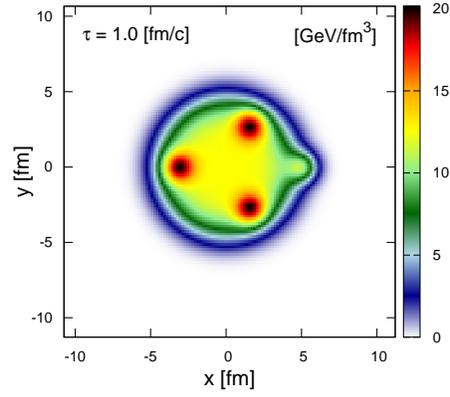}
\vspace*{-0.0cm}
\caption{(Color online) Energy distribution of four tubes placed on top of the average NeXuS IC, the sextupole $\varepsilon_3$ of the distribution is zero.}
\label{fig10}
\end{figure}

\begin{figure}
\begin{tabular}{cc}
\begin{minipage}{200pt}
\centerline{\includegraphics[width=220pt]{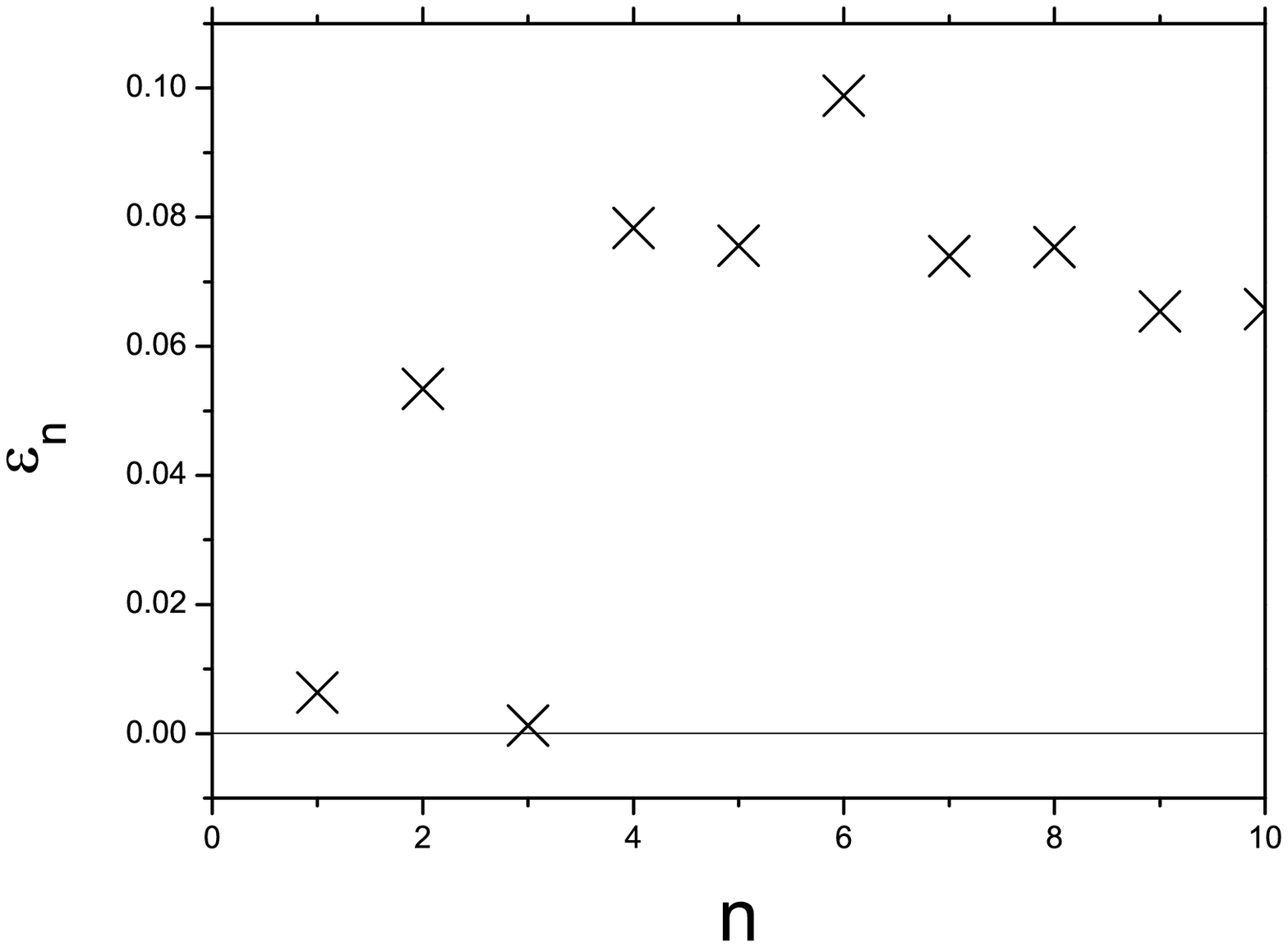}}
\end{minipage}
&
\begin{minipage}{200pt}
\centerline{\includegraphics[width=220pt]{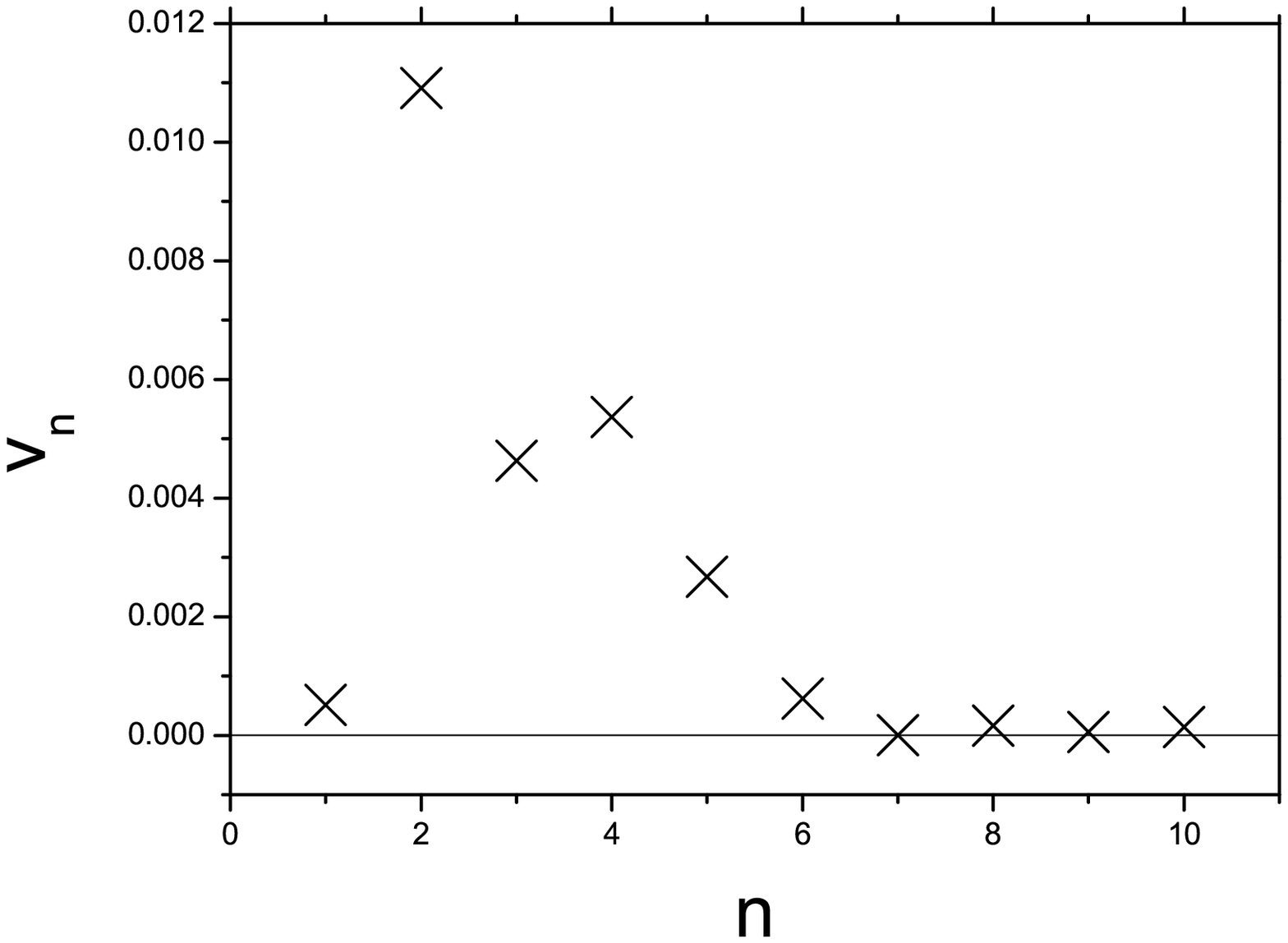}}
\end{minipage}
\\
\end{tabular}
 \caption{Initial eccentricities and calculated flow harmonics of the IC in Fig.10.}
 \label{fig11}
\end{figure}

\end{document}